\documentclass[12pt]{article}
\usepackage{color,epsfig,amsmath,amssymb,graphicx}

\font\srm=cmr8

\def\be{\begin{equation}}
\def\fe{\end{equation}}

\def\spose#1{\hbox to 0pt{#1\hss}}
\def\lta{\mathrel{\spose{\lower 3pt\hbox
{$\mathchar"218$}}\raise 2.0pt\hbox{$\mathchar"13C$}}}  \def\gta{\mathrel
{\spose{\lower 3pt\hbox{$\mathchar"218$}}\raise 2.0pt\hbox{$\mathchar"13E$}}}

\def\Libra{\spose {--} {\cal L}}
\def\Euro{\spose {\lower 2.5pt\hbox{${^{\bf =}}$}}{ C}}

\def\spose#1{\hbox to 0pt{#1\hss}}

\def\sqr#1#2{{\vcenter{\hrule height.4pt\hbox{\vrule width.8pt height#2pt
\kern#1pt\vrule width.8pt}\hrule height.4pt}}}

\definecolor{violet}{rgb}{0.4,0,0.4}
\definecolor{vert}{rgb}{0,0.5,0.0}
\definecolor{navy}{rgb}{0.0,0.0,0.6}
\definecolor{orange}{rgb}{0.8,0.2,0.0}
\definecolor{bleu}{rgb}{0.3,0.0,0.8}
\definecolor{brun}{rgb}{0.6,0.3,0.0}

\def\rms{ {\rm S}}

\def\rmr{ {\color{red}{\rm r}}}
\def\rmf{ {\rm f}}  \def\rmc{ {\rm c}}
\def\rmI{ {\rm I}} 

\def\ses{{\color{vert}s}}
\def\nn{{\color{vert}n}}
\def\qq{{\color{vert}q}}
\def\nc{\nn_{\rmc}}
\def\nf{\nn_{\rmf}}

\def\EU{{\color{brun}\Euro}}
\def\rrho{{\color{brun}\rho}}
\def\TT{{\color{brun}T}}
\def\UU{{\color{brun}U}}
\def\ff{{\color{brun}f}}
\def\lamb{{\color{brun}{\lambda}}}
\def\LP{{\color{brun}{\cal P}}}
\def\PP{{\color{brun}{P}}}
\def\SeS{{\color{brun}{S}}}
\def\Shear{{\color{brun}{\mit \Sigma}}}
\def\LLambda{{\color{brun}\Lambda}}
\def\PPsi{{\color{brun}\Psi}}
\def\PPi{{\color{brun}{\mit\Pi}}}

\def\calQ{{\color{orange}{\cal Q}}}
\def\LmP{{\color{red}{\cal P}}}
\def\calE{{\color{red}{\cal E}}}
\def\Thet{{\color{red}{\Theta}}}
\def\mm{{\color{red}m}} 

\def\hhbar{{\color{red}\hbar}}

\def\ww{{\color{red} w}}
\def\pp{{\color{red}{p}}}

\def\hhbar{{\color{red}{\hbar}}}

\def\uu{{\color{violet}u}} 
\def\vv{{\color{violet}v}}
\def\Vv{{\color{violet}V}}

\def\rr{{\color{bleu}r}} \def\tt{{\color{bleu}t}}

\def\xx{{\color{bleu}x}} \def\eg{{\color{bleu}e}}
\def\ggamma{{\color{bleu}\gamma}}
\def\eeta{{\color{bleu}\eta}}
\def\ddelta{{\color{bleu}\delta}}

\def\gamm{{\color{violet}\gamma}}
\def\sigm{{\color{violet}\varsigma}}
\def\pphi{{\color{violet}\phi}}
\def\varpphi{{\color{violet}\varphi}}
\def\varppi{{\color{red}\varpi}}
\def\ppi{{\color{red}\pi}} 
\def\pperp{{\color{red}\perp}}

\def\mmu{{\color{red}\mu}} 
 
\def\cchi{{\color{red}\chi}}
\def\kkappa{{\color{orange}\kappa}}
\def\cc{{\color{orange}c}}

\begin{document}

\title{\bf Newtonian mechanics of neutron superfluid in elastic
star crust.}

\author { {\bf Brandon Carter \& Elie Chachoua }\\
 \hskip 1 cm\\   \\Observatoire de
Paris, 92195 Meudon, France.}

\date{\it 26 January, 2006}

\maketitle

\vskip 1  cm

{\bf Abstract}  To account for pulsar frequency glitches, it is 
necessary to use a neutron star crust model allowing not only for  
neutron superfluidity but also for  elastic solidity. These
features have been treated separarately in previous treatments of 
crust matter, but are combined here in a unified treatment that is
based on the use of a Lagrangian master functon, so that the 
coherence the system is ensured by the relevant Noether identities.
As well as the model obtained directly from the variation 
principle, the same master function can provide  other 
conservative alternatives, allowing in particular for the effect 
of perfect vortex pinning. It is also shown how such models 
can be generalised to allow for dissipative effects, including
that of imperfect pinning, meaning vortex drag or creep.

\vskip 1  cm

\section{Introduction}

Almost immediately after the discovery of pulsars, and their 
identification as rotating neutron stars, it was recognised
that, to account for their observed frequency glitches, a model 
of the purely fluid type such as is commonly sufficient in stellar 
structure theory would not suffice, and that an adequate
 ``basic picture'' \cite{Horvath04} would require the use of
a more elaborate kind of model allowing for the elasticity 
of the solid crust. In order to do this, a suitable category 
of elastic solid models was developed in which, as well as 
allowing for the effect of General Relativity (which, in the crust 
layers, is significant but not of overwhelming importance) the main 
innovation \cite{CQ72} was to use a treatment based on elastic 
perturbations not with respect to a fully relaxed local state (such 
as will usually be available in the context of terrestrial 
engineering) but with respect to a local state that is only 
conditionally relaxed with respect to perturbations that preserve 
its density (which in a neutron star may be extremely high compared 
with the density at which the solid would be fully relaxed).

The motivation for the effort needed to construct global neutron 
star models \cite{CQ75} of such a purely elastic type was however 
diminished to some extent after it became clear that solidity alone 
could not explain the high rate at which quite large glitches were
actually observed to occur. To account for this, it was generally 
recognised to be necessary to invoke a mechanism involving angular 
momentum transfer from a rather  more rapidly rotating superfluid 
neutron constituent that can flow without resistance through the 
ionic crust material. To describe such a phenomenon at a macroscopic
(local but not global) level, various multiconstituent fluid models 
have been developed over the years, both in a Newtonian framework 
\cite{Mend91} and also (not so much for the minor improvement in 
acuracy that it provides in principle, as because it is for many 
purposes actually easier to work in practise!) in a relativistic 
framework  \cite{CK92,Lan98,Comer03}.

Work on the application, at the global stellar structure level, of 
such multiconstituent fluid models has by now been developed, both in 
a Newtonian framework \cite{Prix02}, and in a relativistic framework 
\cite{Prix05}, to such an extent that it now seems worthwhile to make 
the further step of incorporating the effects of solidity, without 
which the occurence  of the actual glitch phenomenon would not be 
possible. In order to provide what is missing \cite{CC06} in a purely 
fluid descripton of such phenomena, the present article shows how a 
suitably chosen master function $\LLambda$ can be used to set up a 
variational or more general Newtonian model of a neutron superfluid 
flowing within an elastic solid background, using a formalism that 
synthesises the separate descriptions of solidity~\cite{CCC} and 
multiconstituent fluidity~\cite{CC03,CC04} that are already available 
in a form adapted for this purpose. An analogous synthesis in a 
relativistic framework is under preparation elsewhere \cite{CS06}.

\section{Newtonian action formulation}

\subsection{Dynamic variables and background fields}

Our purpose in this Section  is to set up a multipurpose Newtonian 
variational formalism that combines the ones that have recently been 
developed on one hand for the treatment of a multiconstituent fluid 
\cite{CC03,CC04} and on the other hand for a non conducting elastic solid 
\cite{CCC}. The treatment in the present section does not allow for  
dissipative effects, but is nevertheless designed in such a way as to 
facilitate their subsequent inclusion in the Section \ref{Dissip}, using 
the approach that has recently been developed for the multiconstituent 
fluid case \cite{CC05}.

Let us start by recapitulating some basic considerations for the generic 
case,  in a spacetime with arbitrarily chosen coordinates $\xx^\mu$ 
($\mu=0,1,2,3$), of an action density scalar, $\LLambda$ say, whose role 
is to act as a Lagrangian master function governing the evolution of a 
set of active or ``live'' dynamical fields, but whose complete 
specification also involves a set of given -- passive or ``dead'' -- 
background fields.

The most general infinitesimal Eulerian (i.e. fixed point) variation of
such a Lagrangian will be decomposable as the sum of two contributions
in the form
{\be \delta\LLambda=\delta^{_\heartsuit\!}\LLambda +\delta^\ddagger\LLambda
\label{(0)} \fe}
in which $\delta^{_\heartsuit\!}\LLambda$ is the realisable part 
attributable to a physically possible alteration of the configuration of 
the ``live'' dynamical fields, while $\delta^\ddagger\LLambda$ is a virtual 
part arising from mathematically conceivable but (in the context under 
consideration) physically forbidden variations of the ``dead'' background 
fields that have been fixed in advance. In a typical special relativistic 
application the only relevant ``dead'' field might just be the Minkowski 
background metric. What we shall be concerned with here however is the more 
complicated Newtonian case in which, as explained in recent preceding 
work~\cite{CCC}, the necessary set of ``dead'' fields include the  rank-3 
spacemetric $\ggamma^ {\mu\nu}$ and the associated time covector $\tt_\mu$
(which are restricted to be respectively spacelike and timelike in the 
sense of satisfying $\tt_\mu\ggamma^{\mu\nu}=0$) together with  a (Galilean 
gauge fixing) ether vector $\eg^\mu$.  As well as these indispensible 
uniform background fields, the set of the ``dead'' background fields that 
are fixed in advance will be taken here, in the first instance, to include 
the generically non uniform gravitational potential $\pphi$, so that the 
complete background variation contribution will be given by an expression of 
the form
{\be  \delta^\ddagger\LLambda=\frac{\partial\LLambda}
{\partial\ggamma^{\mu\nu}}\delta\ggamma^{\mu\nu}+
\frac {\partial\LLambda}{\partial \tt_\mu}\delta \tt_\mu
+\frac{\partial\LLambda}{\partial \eg^\nu}\delta \eg^\nu+
 \frac{\partial\Lambda}{\partial \pphi}\delta \pphi\, . \label{(1)} \fe}
For more general purposes, the scalar $\pphi$ would need to be promoted 
from being a given background to the status of a ``live'' dynamical
field, which means that the last term in (\ref{(1)}) would be transfered to 
the ``live'' variation contribution $\delta^{_\heartsuit\!}\LLambda$, which
in that case \cite{CC04} would also include a term allowing for the 
dependence of $\LLambda$ not just on $\pphi$ but also on its gradient with 
components $\pphi_{,\mu}$.

It is to be remarked that the construction of many of the relevant 
quantities will involve the rank 3 Euclidean space metric that is induced 
by the choice of the ether gauge vector  $\eg^\mu$ acording to the 
specifications
{\be \eeta_{\mu\nu}\eg^\nu=0\, ,\hskip 1 cm\eeta_{\mu\nu}\ggamma^{\nu\rho}
=\ddelta^\rho_\mu-\eg^\rho\tt_\mu=\eeta^\rho_{\,\mu}\, .\label{5b}\fe}
In the more specialised work on the fluid case \cite{CC04} this uniform 
space metric was denoted by $\ggamma_{\mu\nu}$, but the latter will be used 
here for the variable space metric that is to be defined below.
As noted in our preceding work \cite{CCC}, the induced variation of the 
uniform space metric (\ref{5b}) will be expressible (using round 
brackets to indicate index symmetrisation) by the formula
{\be  \delta\eeta_{ \mu\nu} =-\eeta_{\mu\rho} \eeta_{\nu\sigma}\,\delta
\gamma^{\rho\sigma}-2\tt_{(\mu}\eeta_{\nu)\rho}\,\delta\eg^\rho\, ,\fe}
in which the last term shows the effect just of an infinitesimal Galilean 
transformation, under which $\delta\gamma^{\rho\sigma}$ would simply vanish.

The particular kind~\cite{CC03} of background variation in which we shall 
be interested is one that is simply generated by the action of a 
displacement field $\xi^\nu$ say, so that the ensuing field variations will 
be given just by the negatives of the corresponding Lie derivatives. For 
the background fields that are uniform (in the sense of having vanishing 
covariant derivatives) the resulting variations will be given just by
{\be \delta\ggamma^{\mu\nu}=2\ggamma^{\rho(\mu}\nabla_{\!\rho}\xi^{\nu)}
\hskip 1 cm \delta \tt_\mu=-\tt_\nu\nabla_{\!\mu}\xi^\nu\, ,\hskip 1 cm
\delta \eg^\nu=\eg^\mu\nabla_{\!\mu}\xi^\nu\, ,\label{(2)}\fe}
while for the gravitational potential we shall simply have
{\be \delta\pphi=-\xi^\nu\nabla_{\!\nu}\pphi\, .\label{(3)}\fe}

In so far as the ``live'' contribution is concerned, a further subdivision
arises in the cases such as those of interest here, which are characterised
by a constrained variation principle, meaning one whereby the ``on shell'' 
evolution condition is that of invariance of the relevant action integral 
with respect to a compactly supported perturbation of the dynamical fields 
that is not entirely arbitrary but constrained by an appropriate 
admissibility condition. In the applications under consideration, the 
relevant admissibility condition will be interpretable as the requirement 
for the change to represent a ``natural'' variation of the same given 
physical system, whereas more general changes would represent a replacement
of the system by a slightly different system within the same category.
In such a case the ``live'' field contribution will be decomposible (though 
not necessarily in a unique manner) as a sum of the form
{\be \delta^{_\heartsuit\!}\LLambda=\delta^\natural\LLambda+\delta^\sharp
\LLambda \label{(4)}\fe}
in which $\delta^\sharp\LLambda$ denotes a part the would be inadmissible 
for the purpose of application of the variational principle, whereas 
$\delta^\natural\LLambda$ is a ``natural'' variation that would be allowed 
for this purpose. This means that for any (unperturbed) configuration that 
is ``on shell'' -- in the sense of satisfying the dynamical evolution 
equations provided by the constrained variation principle -- a generic 
admissible variation must satisfy  condition
{\be \delta^\natural\LLambda\cong 0\, ,\label{(5)}\fe}
using the symbol $\cong$ to indicate equivalence modulo the addition of 
a divergence (which, by Green's theorem, will give no contribution to the 
action integral from a variation that is compactly supported).

In the kind of medium with which we are concerned, the constrained 
variables will be a set of current 4-vectors $\nn_{\rm _X}^{\,\mu}$
with ``chemical'' index label {\srm X} (which, in the kind of astrophysical
aplications we have in mind, might include the separate neutronic and 
protonic contributions to the conserved total baryonic current) while the 
unconstrained fields will consist of a  triplet of scalar fields 
$\qq^{\rm _A}$ ({\srm A} =1,2,3) that are interpretable \cite{CCC} as local 
coordinates on a 3-dimensional material base manifold, and from which, 
together with their gradient components $\qq^{\rm _A}_{\, ,\mu}$, an 
associated particle number (which might count ionic nuclei or crystaline 
solid lattice points in the relevant microscopic substructure) density 
current, $\nn_{\rm I}^{\,\nu}$ say, can be constructed in the manner 
prescribed below. The corresponding ``live'' part of the generic action 
variation will thus be expressible in the form
{\be \delta^{_\heartsuit\!}\LLambda=\Sigma_{\rm _X}
\ppi^{\rm _X}_{\ \mu}\delta \nn_{\rm _X}^{\,\mu}
+\LmP^\nu_{\,\rm_A}\delta \qq^{\rm _A}_{\, ,\nu}
+\frac{\partial \LLambda}{\partial \qq^{\rm _A}}\delta\qq^{\rm _A}
\, ,\label{(15)}\fe}
in which, for each separate current, the partial derivative
$\ppi^{\rm _X}_{\ \mu}=\partial\LLambda/\partial \nn_{\rm _X}^{\,\mu}$
is interpretable as the corresponding 4-momentum per particle. 

It is to be observed that the partial derivative $\LmP^\nu_{\,\rm_A}=
{\partial \LLambda}/{\partial \qq^{\rm _A}_{\, ,\nu}}$ has the
noteworthy property of being tensorial not just with respect to
transformations of the space time coordinates $\xx^\nu$ but also with 
respect to transformations $\qq^{\rm_A}\rightarrow \tilde \qq{^{\rm_A}}$
of the material coordinates, $\qq^{\rm_A}$ whose effect will simply
be given by
{\be \qq^{\rm_A}\mapsto \tilde \qq{^{\rm_A}}\hskip 1 cm
\Rightarrow \hskip 1 cm\LmP^\nu_{\,\rm_A}\mapsto 
\LmP^\nu_{\,\rm_B}\,\frac{\partial \qq^{\rm_B}}{\partial \tilde
 \qq{^{\rm _A}}}\, ,\label{15a}\fe}
This good behaviour contrasts with the comportment of the remaining 
partial derivative, ${\partial\LLambda}/{\partial\qq^{\rm _A}}$,
for which the corresponding transformation has the non-tensorial form  
{\be\frac{\partial\LLambda}{\partial\qq^{\rm _A}}\mapsto\frac
{\partial\LLambda}{\partial\qq^{\rm _B}}\,\frac{\partial \qq^{\rm_B}}
{\partial \tilde \qq{^{\rm _A}}} +\LmP^\nu_{\,\rm_C}\,
\tilde\qq{^{\rm _B}_{,\nu}}\, \frac{\partial^2\qq^{\rm_C}}
{\partial \tilde \qq{^{\rm _B}}\partial \tilde \qq{^{\rm _A}}}
\, .\label{15b}\fe}

For each separate current $\nn_{\rm _X}^{\,\mu}$, the admissible variations 
(as generated~\cite{CC03} by worldline displacements) will be specifiable 
by a corresponding displacement vector field, $\xi_{\rm _X}^{\,\mu}$
say,  by the formula
{\be \delta^\natural \nn_{\rm _X}^{\,\mu}=-\vec\xi_{\rm _X}
\Libra\, \nn_{\rm _X}^{\,\mu}-\nn_{\rm _X}^{\,\mu}\nabla_{\!\nu\,}
\xi_{\rm _X}^{\,\nu}\, ,\label{(6)}\fe}
in which the Lie derivative is just the commutator
{\be  \vec\xi_{\rm _X}\Libra\, \nn_{\rm _X}^{\,\mu}=\xi_{\rm _X}^{\,\nu}
\nabla_{\!\nu\,}\nn_{\rm _X}^{\,\mu}-\nn_{\rm _X}^{\,\nu}
\nabla_{\!\nu\,}  \xi_{\rm _X}^{\,\mu}\, .\label{(7)}\fe}

The variations of the material base coordinate fields $\qq^{\rm _A}$
will similarly be expressible in terms of their own displacement vector
field $\xi^\mu$, by an expression of the corresponding form
{\be \delta^\natural \qq^{\rm _A}=-\qq^{\rm _A}_{\, ,\mu}\xi^{\,\mu}
\, ,\label{(8)}\fe}
but in this case it is evident that no restriction is entailed, so that
without loss of generality we can take $\delta^\sharp \qq^{\rm A}=0$,
which means that the inadmissible part (if any) of a ``live'' variation
will reduce simply to
{\be \delta^\sharp\LLambda=\Sigma_{\rm _X} \ppi^{\rm _X}_{\  \nu}
\,\delta^\sharp\nn_{\rm _X}^{\,\nu}\, .\label{(8a)}\fe}
It can be seen (using integration by parts) that, in terms of the
displacement fields introduced by (\ref{(6)}) and (\ref{(8)}), the generic 
admissible variation for the current carrying medium will be expressible 
--modulo a variationally irrelevant divergence term -- in the form
{\be  \delta^\natural\LLambda\cong-\ff^{\rms}_{\, \nu}\,\xi^{\nu}
-\Sigma_{\rm _X}\ff^{\rm _X}_{\, \nu}\,\xi_{\rm _X}^{\, \nu} 
\, ,\label{(9)}\fe}
in which, for each value of {\srm X},  the covectorial coefficient 
$\ff^{\rm _X}_{\, \nu}$ will be interpretable as the non-gravitational 
force density acting on the corresponding separate constituent, while 
the extra  coefficient $\ff^{\rms}_{\,\nu}$ is the supplementary force 
density due, as discussed below, to stratification or solid elasticity,
that acts on the underlying ionic (crystalline or glass like) lattice 
structure. 

If we postulate that the system should obey the constrained variational 
principle that is expressible as the imposition of the ``on shell'' condition 
(\ref{(5)}), then it is evident from (\ref{(9)}) that the ensuing dynamical 
field equations will be expressible simply as the separate vanishing of each 
one of these force densities, i.e.  as the requirement that we should have 
$\ff^{\rms}_{\, \nu}=0$ and (for each value of {\srm X}) 
$\ff^{\rm _X}_{\, \nu}=0$. We shall however be mainly concerned with other
possibilities, in particular with cases in which some of the constituents,  labelled 
{\srm X}=c and {\srm X}$=\emptyset$, are ``confined'' in the sense of being 
subject to a convection condition of the form
{\be \nn_{\rmc}^{\,\nu}\qq^{\rm _A}_{\, ,\nu}=0 \, ,\hskip 1 cm
 \nn_\emptyset^{\,\nu}\qq^{\rm _A}_{\, ,\nu}=0
\, , \label{11a}\fe}
which means that they have to move with the underlying ionic flow.
When the application of the variation principle is subject to
the corresponding convective constraints
{\be \xi_{\,\rmc}^{\, \nu}= \xi_\emptyset^{\, \nu}=
\xi^{\nu}\, ,\label{11b}\fe}
the ensuing system of dynamical field equations will no longer entail
the separate vanishing of  $\ff^{\rms}_{\, \nu} , $ 
$\ff^\emptyset_{\, \nu}$ and $\ff^{\rmc}_{\,\nu}$,
but only of  the amalgamated ionic force density $\ff^{\rmI}_{\, \nu}$ 
that is defined as their sum,
{\be \ff^{\rmI}_{\, \nu}= \ff^{\rms}_{\,\nu}+\ff^{\rmc}_\nu+
\ff^\emptyset_\nu\, ,\label{12}\fe}
which is interpretable as the net force density acting on the integrated 
system consisting of the ions in combination with the convected constituent.

\subsection{Noether identity for Newtonian covariance}

As in the preceding work~\cite{CC03}, let us now consider the Noether type
identity that is obtained (regardless of whether or not the variational
field equations are satisfied) by taking the variations of all the relevant
fields, ``live'' as well as ``dead'' to be given by the action of the
same displacement field $\xi^\nu$ whose effect on the background fields 
was given by (\ref{(2)}) and (\ref{(3)}), so that the resulting effect on 
the Lagrangian scalar will be given just by the corresponding displacement
variation
{\be \delta\LLambda=-\xi^\nu\nabla_{\!\nu}\LLambda\, .\label{(10)}\fe}
For the unconstrained fields $\qq^{\rm _A}$ the effect of such a 
displacement will be given, according to  (\ref{(8)}), simply by
{\be \delta\qq^{\rm _A}= -\vec\xi\Libra \qq^{\rm _A} =
\delta^\natural \qq^{\rm _A} \, .\label{(11)}\fe}
On the other hand, for the constrained variables $\nn_{\rm _X}^{\,\nu}$ ,it can be seen 
by setting  $\xi_{\rm _X}^{\, \nu}$ to $\xi^\nu$ in (\ref{(6)})  that we 
shall have
{\be \delta\nn_{\rm _X}^{\,\nu}=-\vec{\xi}\Libra \nn_{\rm _X}^{\,\nu}
=\delta^\natural
\nn_{\rm _X}^{\,\nu}+\delta^\sharp\nn_{\rm _X}^{\,\nu} \, ,\label{(12)}\fe}
in which the extra ``unnatural'' (variationally inadmissible) contribution  
will be given by
{\be \delta^\sharp\nn_{\rm _X}^{\,\nu}=\nn_{\rm _X}^{\,\nu}
\nabla_{\!\mu}\xi^\mu\, .\label{(13)}\fe}
It can thereby be seen from (\ref{(0)}) using (\ref{(9)}) and (\ref{(10)}) 
that, as an identity which will hold, modulo a divergence, for any 
dynamical field configuration whether or not it is ``on shell'', we shall 
have a relation of the form
{\be \xi^\nu\left(\ff^{\rms}_{\,\nu}+\Sigma_{\rm _X}
\ff^{\rm _X}_{\, \nu} -\nabla_{\!\nu}\LLambda\right)
\cong \delta^\sharp\LLambda+\delta^\ddagger\LLambda\, .\label{(14)}\fe}
We now evaluate the terms on the right by substitution of (\ref{(13)}) 
in (\ref{(8a)}) and by substitution of the formulae (\ref{(2)}) and 
(\ref{(3)}) in (\ref{(1)}) for an arbitrary displacement field $\xi^\nu$. 
The identity (\ref{(14)}) can thereby be rewritten more explicitly in the 
form
{\be \xi^\nu\left(\ff^{\rms}_{\,\nu}+\Sigma_{\rm _X}\ff^{\rm _X}_{\, \nu} 
-\rrho\nabla_{\!\nu}\pphi\right)\cong -\TT^\mu_{\ \nu}\nabla_{\!\mu}\xi^\nu
\cong\xi^\nu\nabla_{\!\mu}\TT^\mu_{\ \nu}\, ,\label{(16)}\fe}
in which the effective (passive) gravitational mass density can be read
out simply as
{\be \rrho=-\frac{\partial\LLambda}{\partial\pphi}\, ,\label{(17)}\fe}
while the corresponding (gravitationally passive) geometric stress 
momentum energy density tensor can be read out as
{\be \TT^\mu_{\ \nu}=\left(\LLambda-\Sigma_{\rm _X}
\ppi^{\rm _X}_{\ \rho}\nn_{\rm _X}^{\,\rho}\right)\delta^\mu_{\,\nu}
 - 2\ggamma^{\mu\rho}\frac{\partial\LLambda}{\partial\ggamma^{\rho\nu}}+\frac
{\partial\LLambda}{\partial \tt_\mu} \tt_\nu- \eg^\mu\frac{\partial \LLambda}
{\partial \eg^\nu}\, . \label{(18)}\fe}
Since the arbitrary field $\xi^\nu$ can be taken to be non zero only
in the immediate neighbourhood of any chosen point, the Noether identity
(\ref{(16)}) implies that, at each point, this stress energy tensor
will satisfy a divergence identity of the simple form
{\be \nabla_{\!\mu}\TT^\mu_{\ \nu}=\ff^{\rms}_{\,\nu}
+\Sigma_{\rm _X}\ff^{\rm _X}_{\, \nu} -\rrho\nabla_{\!\nu}\pphi
\, .\label{(19)}\fe}
In the ``on shell'' case for which the variational evolution equations are
satisfied, the non gravitational force density contributions on the right
of (\ref{(19)}) will drop out, leaving just the final term on the right 
which represents the gravitational force density which will always be 
present since the gravitational potential $\pphi$ is treated as a background field.

The tensor $\TT^\mu_{\ \nu}$ can be employed in the usual way for  
construction of the energy flux vector $\UU^\nu$ and the energy denstity 
$\UU$ according to the prescriptions
{\be \UU^\mu=-\TT^\mu_{\ \nu}\eg^\nu\, ,\hskip 1 cm \UU=\tt_\mu\UU^\mu
\, ,\label{20}\fe}
while the corresponding space momentum flux $\PPi^{\mu\nu}$ and space 
momentum density $\PPi^\mu$ will be given by the prescriptions
{\be \PPi^{\mu\nu}= \TT^\mu_{\ \sigma}\ggamma^{\sigma\nu} \, \hskip 1 cm 
\PPi^\nu =\tt_\mu  \PPi^{\mu\nu}\, ,\label{20a}\fe}
which can then be used to determine a pressure tensor $\PP^{\mu\nu}$
that is specified by the decomposition
{\be \PPi^{\mu\nu}=\PP^{\mu\nu}+\eg^\mu\PPi^\nu\, .\label{20e}\fe}
( It is to be remarked, as a minor caveat, that the systematic
notation scheme previously developed for cases in which all 
the constituents were of purely fluid type \cite{CC04} deviated slightly
from the -- even more systematic -- scheme used here, not only by using 
$\ggamma_{\mu\nu}$ for what is denoted here by $\eeta_{\mu\nu}$, but also 
by using $\PP^\mu_{\ \nu}$ just as an abbreviation for  
$\TT^{\,\mu}_{\!_{\rm int} \sigma}$.) It can be seen from the Noetherian 
construction (\ref{20}) that the pressure tensor that is specified in this 
way will automatically be symmetric and strictly spacelike,
{\be \PP^{\mu\nu}=\PP^{\nu\mu}\, ,\hskip 1 cm \PP^{\mu\nu}\tt_\nu=0
\, ,\label{20c}\fe}
since it will be given explicitly by 
{\be  \PP^{\mu\nu}=\left(\LLambda-\Sigma_{\rm _X}
\ppi^{\rm _X}_{\  \sigma}\nn_{\rm _X}^{\,\sigma}\right)\ggamma^{\mu\nu}
- 2\ggamma^{\mu\rho}\frac{\partial\LLambda}{\partial\ggamma^{\rho\sigma}}
\ggamma^{\sigma\nu}\, .\label{20d}\fe}
It can also be seen that the 3-momentum density will contain no 
contribution from any part of the Lagrangian that 
is independent of the Galilean frame, as is the case for the 
purely internal action contribution $\LLambda_{\rm_{int}}$ that will be 
discussed below: we shall simply have
{\be\PPi^\nu =-\frac {\partial \LLambda}{\partial \eg^\sigma}
\ggamma^{\sigma\nu}\hskip 1 cm \Rightarrow \hskip 1 cm 
\PPi_{\!\rm_{int}}^{\,\nu}= 0.\label{20b}\fe}
The corresponding formula for the scalar energy density is
{\be \UU=\Sigma_{\rm _X}\ppi^{\rm _X}_{\  \sigma}\nn_{\rm _X}^{\,\sigma}
-\LLambda-\tt_\sigma\frac{\partial \LLambda}{\partial \tt_\sigma}+
\eg^\sigma\frac{\partial \LLambda}{\partial \eg^\sigma}\, .\fe}

\subsection{Canonical formulation}

If, instead of working it out in terms of force densities modulo a
divergence as in (\ref{(14)}) and (\ref{(16)}), we insert the displacement 
contributions (\ref{(11)}) and (\ref{(12)}) to the ``live'' variation 
(\ref{(15)}) directly in (\ref{(0)}) we get an identity in which the 
coefficients of the locally adjustable fields $\xi^\mu$ and 
$\nabla_{\!\mu}\xi^\nu$ must vanish separately. The former of these 
conditions reduces to a triviality, but the latter provides a relation
showing that the geometrically defined stress energy tensor (\ref{(19)}) 
can be rewritten in the equivalent canonical form
{\be \TT^\mu_{\ \nu}=\LLambda\delta^\mu_{\ \nu}+\Sigma_{\rm _X}\left(
\ppi^{\rm _X}_{\  \nu}\nn_{\rm _X}^{\,\mu}-
\ppi^{\rm _X}_{\  \rho}\nn_{\rm _X}^{\,\rho}\,\delta^\mu_{\,\nu}\right)
-\LP^\mu_{\ \nu}\, ,\label{(21)}\fe}
in which the stress contribution at the end is given in terms
of the coefficient introduced in (\ref{15a}) by
{\be \LP^\mu_{\ \nu}=\LmP^\mu_{\,\rm _A}\qq^{\rm _A}_{\, ,\nu}
\, .\label{21a}\fe}

The ``live'' variation formula (\ref{(15)}) also provides what is needed 
for the derivation of the corresponding expressions for the force densities
introduced in (\ref{(9)}) which for the fluid constituents will have the 
form that is already familiar from the preceding work~\cite{CC03}. In 
terms of the corresponding generalised vorticity 2-forms defined, using 
square brackets for index antisymmetrisation, by
{\be \varppi^{\rm _X}_{\,\mu\nu}=2\nabla_{\![\mu}
\ppi^{\rm _X}\!{_{\!\nu]}}\, ,\label{(22a)}\fe}
these non-gravitational current force densities will be given by
{\be \ff^{\rm _X}_{\, \nu}= \nn_{\rm _X}^{\,\mu} 
\varppi^{\rm _X}_{\,\mu\nu}+\ppi^{\rm _X}_{\ \nu} \nabla_{\!\mu}
\nn_{\rm _X}^{\,\mu}\, ,\label{(22)}\fe}
while the non-gravitational force density acting on the underlying atomic 
structure of the medium will be given by the prescription
{\be \ff^{\rms}_{\,\mu}=\frac{\delta \LLambda}{ \delta \qq^{\rm _A}}
\qq^{\rm _A}_{\, ,\mu}\, ,\label{(23)}\fe}
in which the Eulerian derivative is given by a prescription of the
usual form
{\be \frac{\delta \LLambda}{ \delta \qq^{\rm _A}}=\frac{\partial\LLambda}
{\partial  \qq^{\rm _A}}-\nabla_{\!\nu}\LmP^\nu_{\,\rm _A}
\, ,\hskip 1 cm \LmP^\nu_{\,\rm _A}=\frac{\partial\LLambda}
{\partial  \qq^{\rm _A}_{\, ,\nu}} \, .\label{(24)}\fe}
It is important to notice that the non-tensorial base coordinate
transformation property  (\ref{15b}) of the first term in this formula 
will be cancelled by that of the second, so that the effect on the Eulerian
derivative of a change of the material base coordinates will be
given simply by
{\be \qq^{\rm_A}\mapsto \tilde \qq{^{\rm_A}}\hskip 1 cm \Rightarrow 
\hskip 1 cm \frac{\delta\LLambda}{\delta \qq^{\rm_A}}
\mapsto  \frac{\delta\LLambda}{\delta \qq^{\rm_B}}
\,\frac{\partial \qq^{\rm_B}}{\partial \tilde \qq{^{\rm _A}}}
\label{24a}\, ,\fe}
which shows that the (stratification and solid elasticity) force density 
(\ref{(23)}) is invariant with respect to such transformations, an 
important property that is not so obvious from its detailed expression
{\be \ff^{\rms}_{\,\mu}=\frac{\partial \LLambda}{ \partial \qq^{\rm _A}}
\qq^{\rm _A}_{\, ,\mu}+ \LmP^\nu_{\,\rm _A} \nabla_{\!\mu}
\qq^{\rm _A}_{\, ,\nu}-\nabla_{\!\nu}\left( \LmP^\nu_{\,\rm _A}
\qq^{\rm _A}_{\, ,\mu}\right)\, ,\label{(25)}\fe}
in which only the final term is separately invariant.

It is important to observe that, regardless of what -- variational or
other -- dynamical field equations may be imposed, the force density 
(\ref{(25)}) will never be able to do any work on the medium since, as an 
obvious consequence of (\ref{(23)}) it must automatically satisfy the 
further identity 
{\be \ff^{\rms}_{\,\nu} \uu^\nu=0\, ,\label{(26)}\fe}
where $\uu^\mu$ is the unit 4-velocity of the medium as specified by the 
defining conditions
{\be \qq^{\rm _A}_{\, ,\nu} \uu^\nu=0\, ,\hskip 1 cm \tt_\nu \uu^\nu=1
\, .\label{(27)}\fe}

If the variation principle (\ref{(5)}) is imposed, so that -- as remarked 
above -- by (\ref{(9)}) the separate force densities $\ff^{\rms}_{\,\mu}$ 
and $\ff^{\rm _X}_{\, \mu}$ will all have to vanish, then it evidently 
follows from the Noether identity (\ref{(19)}) that the divergence 
condition 
{\be \nabla_{\!\mu}\TT^\mu_{\ \nu}=-\rrho\nabla_{\!\nu}\pphi 
\, ,\label{(28)}\fe}
must hold. This condition is interpretable as an  energy-momentum
balance equation that must be satisfied whenever the system is not
subject to any non-gravitational external forces. So long as we are
concerned only with a system that is isolated in this sense, so that the 
energy-momentum balance condition (\ref{(28)}) is satisfied, it follows 
from (\ref{(19)}) that, even for a  more general model admitting the 
presence of dissipative or other internal forces, these must be be 
specified in such a way as to satisfy the total force balance condition
{\be \ff^{\rms}_{\,\nu}+\Sigma_{\rm _X}\ff^{\rm _X}_{\, \nu}=0
\, ,\label{(29)}\fe} 
in order for the model to be self consistent. More particularly,
in view of (\ref{(26)}) it can be seen that the fluid force densities
by themselves will have to satisfy the condition
{\be \uu^\mu\Sigma_{\rm _X}\ff^{\rm _X}_{\, \mu}=0\, ,\label{(30)}\fe} 
expressing energy balance in the local rest frame of the underlying
medium.

\section{Application to superfluid neutron conduction in convective solid}

\subsection{Implementation of the confinement constraint}

In the appendix of the third part \cite{CC05} of the preceding series 
of articles on multiconstituent fluid models in a Newtonian framework, 
an account was provided of the kind of application that provided the 
main motivation for this work, namely a model in which the chemical 
variable {\srm X} runs just three values, which are {\srm X}=f and 
{\srm X}= c say, labelling a ``confined'' baryon current 
$\nn_{\rm c}^{\,\nu}$  and a  ``free'' 
baryon current  $\nn_{\rm f}^{\,\nu}$, 
together with a third value, {\srm X}=$\emptyset$ say, labelling an 
entropy current $\ses^\nu=\ses \uu_\emptyset^{\,\nu} , $ which 
(because of the relatively low temperatures involved) will play a 
relatively minor role in the pulsar applications we have in mind.  
The new feature in the present treatment is allowance for the solidity
that (again because of the relatively low temperatures involved) will 
characterise the ``confined''constituent, whose flow 
will therefore be constrained to be aligned with the flow vector
$\uu^\nu$ of the solid structure as introduced above, so that
it will have the form
{\be \nn_{\rm c}^{\,\nu}=\nn_{\rm c}\, u^{\,\nu}\, .\label{(31)}\fe}
As in the previous treatment \cite{CC05} we shall restrict our 
attention here to the case in which the (in any case relatively 
unimportant) entropy current is convected with the solid structure, so 
that the thermal rest frame will be identifiable with that of the 
solid, $\uu_\emptyset^{\,\nu}=\uu^\nu$, which means that the entropy
current will be constrained to have a form analogue to (\ref{(31)}),
namely
{\be \ses^{\,\nu}=\ses\, u^{\,\nu}\, .\label{(31a)}\fe}

The part of the baryon current that is confined according to (\ref{(31)})
(meaning that has to be comoving with the underlying atomic or crystal 
structure of the medium) will add up with the ``free'' part to give a 
conserved total baryon current
{\be\nn_{\rm b}^{\,\nu}= \nn_{\rm c}^{\,\nu} +  \nn_{\rm f}^{\,\nu}\, ,
\hskip 1 cm \nabla_{\!\nu}\nn_{\rm b}^{\,\nu}=0 \, . \label{(32)}\fe}

It can be seen to follow from this current conservation requirement 
that, in terms of the ``free'' current vorticity 2-form 
$\varpi^{\rmf}_{\,\mu\nu} ,$ the energy conservation identity 
(\ref{(30)}) will reduce to the form
{\be \uu^\mu \varppi^{\rm f}_{\,\mu\nu}\nn_{\rm f}^{\,\nu}=
({\calE}^{\rm c} - {\calE}^{\rm f})\nabla_{\!\nu}
\nn_{\rm f}^{\, \nu}-\Thet\nabla_{\!\nu}\ses^{\nu} 
\, ,\label{(33)}\fe}
in which the relevant rest-frame particle energies,  ${\calE}^{\rm c}$
and  ${\calE}^{\rm f}$,  and the temperature, ${\calE}^\emptyset=\Thet$ 
say, are given in terms of the corresponding 4-momenta, as introduced in 
(\ref{(15)}) , by the definitions
{\be {\calE}^{\rm c}=- \uu^\mu\,\ppi^{\rm c}_{\ \mu} \, ,\hskip 1 cm
 {\calE}^{\rm f}=- \uu^\mu\,\ppi^{\rm f}_{\ \mu} \, ,\hskip 1 cm
\Thet=- \uu^\mu\,\ppi^\emptyset_{\ \mu}
\, .\label{(34)}\fe}

It is to be noticed that these definitions (\ref{(34)}) are not affected 
by the freedom to adjust the specifications of the confined particle 
4-momentum covector $ \ppi^{\rm c}_{\ \mu}$ and the thermal 4-momentum 
covector $\ppi^\emptyset_{\ \mu}$ due to the constraints (\ref{(31)}) 
and  (\ref{(31a)}) on the corresponding currents $\nn_{\rm c}^{\,\nu}$ 
and $\ses^\nu$, of which the former can be seen to have the form
(\ref{11a}) which evidently means that the specification
of $\qq^{\rm _A}_{\, ,\nu}$ in (\ref{(15)}) will also be ambiguous.
These ambiguities will however need to be resolved in order for us
to be able to proceed in a well defined manner. To obtain a
formalism that matches smoothly with what has already been set up 
for the fluid limit \cite{CC05}, we need to remove the ambiguity 
in  $\ppi^{\rm c}_{\ \mu}$ by the imposition on 
$\ppi^\emptyset_{\ \mu}$ and $\qq^{\rm _A}_{\, ,\nu}$ of appropriate 
conditions, which can be taken to be that the former should only have 
no space component, which means that will be given in terms of the 
temperature $\Thet$ by 
{\be \ppi^\emptyset_{\ \mu}=- \Thet\, \tt_\mu\, ,\label{(34b)}\fe}
while the latter should have only space components, meaning that 
it will satisfy the condition
{\be  \LmP^\nu_{\,\rm_A}\, \tt_\nu=0\, .\label{(34a)}\fe}
The stress energy tensor (\ref{(21)}) will thereby be obtained 
in the form
{\be \TT^\mu_{\ \nu}=\left(\LLambda+\Thet\ses-\ppi{^{\rmf}_{\,  \rho}}
\nn{_{\rmf}^{\,\rho}}-\ppi{^{\rmc}_{\,  \rho}}
\nn{_{\rmc}^{\,\rho}}\right)\delta^\mu_{\ \nu}-\Thet\ses^\mu\tt_\nu+
\ppi{^{\rmf}_{\,  \nu}}\nn_{\rmf}^{\,\mu}+
\ppi{^{\rmc}_{\,  \nu}}\nn{_{\rmc}^{\,\mu}}
-\LmP{^\mu_{\,\rm _A}}\qq^{\rm _A}_{\, ,\nu}\, .\label{46a}\fe}
(Instead of the convention (\ref{(34a)}), an obvious alternative ansatz 
would be to use the analogue of (\ref{(34b)}) to the effect that  
$\ppi^{\rm c}_{\ \mu}$ should have no space component, so that it would 
take the form $\ppi^{\rm c}_{\ \mu}=-{\calE}^{\rm c} \, t_{\mu}$. Such 
an alternative choice would be adequate for the specialised purpose
of matching the formalism used for the non-conducting solid limit case 
\cite{CCC}, but not for broader objective of consistency with the 
formalism developed for the multiconstituent fluid limit case 
\cite{CC03,CC04,CC05}.)

It is computationally convenient and physically natural to take the material
base space to be endowed with a measure form say that is specifiable in 
terms of  antisymmetric components, $\nn_{\rm I_{ABC}}$ say, that are 
fixed in the sense of depending only on the $\qq^{\rm_A}$, and that will 
determine a corresponding scalar space-time field $\nn_{\rm I}$ by the 
determinant formula
{\be \nn_{\rm I}^{\,2}=\frac{1}{3!} \nn_{\rm I_{ABC}} \nn_{\rm I_{DEF}}\,
\gamm^{_{AD}}\gamm^{_{BE}}\gamma^{_{CF}} \, ,\label{55}\fe}
in which $\gamm^{_{AB}}$ denotes the induced  base space  metric 
{\be \gamm^{_{AB}}=\ggamma^{\mu\nu}\qq^{\rm_A}_{\, ,\mu}
\qq^{\rm_B}_{\, ,\nu} \, , \label{55a}\fe}
whose components will be time dependent, unlike those of the measure 
$\nn_{\rm I_{ABC}}$. For purposes of physical interpretation it will 
usually be convenient to take this measure to represent the number density 
of ionic nuclei, so that the confined baryon number density 
$\nn_\rmc=\tt_\nu\nn_\rmc^{\, \nu}$ will be given by
{\be \nn_{\rm c}={A_\rmc}\nn_{\rm I}\, ,\label{56}\fe}
where ${A_\rmc}$ is the atomic number, meaning the number of confined 
baryons (protons plus confined neutrons) per nucleus. It is to be observed 
that, whereas the number current $\nn_{\rm c}^{\,\nu}$ will fail to be 
conserved in the chemical equilibrium case characterised by (\ref{(42)}), 
the formalism is such that the corresponding ionic number current
{\be \nn_{\rm I}^{\,\nu}=\nn_{\rm I}\, \uu^\mu\fe}
must  automatically satisfy the conservation law
{\be \nabla_{\!\nu} \nn_{\rm I}^{\,\nu}=0\label{57}\fe}
as a mathematical identity.

\subsection{Chemical gauge adjustments}
\label{chemadj}
For particular physical applications, the specification of the part of the 
baryon current that is to be treated as ``confined''will be to some extent 
dependent on the timescales of the processes under consideration as compared 
with the lifetimes for quantum tunnelling through the confining 
barriers containing the nuclei. For relatively rapid processes the fraction 
of the relevant quantum states that should be considered to be confined may 
be subject to an increase, so the corresponding current, $\tilde\nc^\nu$ 
say, of baryons that are effectively confined will be somewhat larger than 
the value, $\nc^\nu$ say, that would be appropriate for processes occuring 
over longer timescales. It is therefore of interest to consider the effect 
of such a chemical basis adjustment, as given by a transformation of the 
form
{\be \tilde\nc{^\nu}=(1+\epsilon)\nc^\nu\, ,\hskip 1 cm
\tilde\nf{^\nu}=\nf^\nu-\epsilon\nc^\nu\, ,\label{44}\fe}
for some suitably specified dimensionless adjustment parameter 
$\epsilon$. So long as $\epsilon$ is just a constant, it is 
immediately apparent that such a transformation will leave the
``free'' 4-momentum covector invariant, meaning that we shall have
{\be \ppi^\rmf_{\,\nu}=\tilde\ppi{^\rmf_{\,\nu}} \, ,\label{45}\fe}
and it is also easy to see that in terms of the new variables the stress 
energy tensor(\ref{46a}) will also be expressible in the same canonical 
form as before, 
{\be \TT^\mu_{\ \nu}=\tilde \TT{^\mu_{\ \nu}}\, .\label{46}\fe}

As in the fluid case \cite{CCH05} this covariance property of the 
canonical stress energy tensor is not restricted to transformations for 
which $\epsilon$ is constant, but can be seen to hold whenever 
$\epsilon$ is specified as a function just of the atomic number 
$A_\rmc=\nn_\rmc/\nn_\rmI$, and of the material position coordinates 
$\qq^{\rm_A}$. Since it can be seen from (\ref{55}) that the variation of 
the ionic number will have a ``live'' (fixed background) part expressible 
in the form
{\be\delta^{_\heartsuit\!}\nn_{\rmI}=\nn_{\rm I}\, \gamm_{\rm_{AB}}
\qq^{\rm _B}_{\, ,\mu} \gamma^{\mu\nu}\delta\qq^{\rm _A}_{\, ,\nu}
+\frac{\partial\nn_\rmI}{\partial  \qq^{\rm _A}}\delta\qq^{\rm_A}
\, ,\label{62}\fe}
it follows that the corresponding variation of the adjustment parameter 
will have the form
{\be\delta^{_\heartsuit\!}\epsilon=\frac{\partial\epsilon}
{\partial A_\rmc}\left(\frac{\delta^{_\heartsuit\!} \nc}{\nn_\rmI}
-A_\rmc\gamm_{\rm_{AB}} \qq^{\rm _B}_{\, ,\mu} 
\gamma^{\mu\nu}\delta\qq^{\rm _A}_{\, ,\nu} \right)
+\left(\frac{\partial\epsilon}{\partial\qq^{\rm _A}}
-\frac{\partial \epsilon}{\partial A_\rmc}\frac{A_\rmc}{\nn_\rmI}
\frac{\partial \nn_\rmI}{\partial \qq^{\rm _A}} 
\right)\delta\qq^{\rm_A}\, .\label{47}\fe}
Since we have $\delta^{_\heartsuit\!} \nc=\tt_\nu\,\delta\nn_\rmc^\nu$
it follows that unlike the free particle momentum, which is subject to
the chemical invariance condition (\ref{45}), the confined particle
momentum covector will undergo a non-trivial chemical adjustment 
given by
{\be \ppi^\rmc_{\,\nu}=\tilde\ppi{^\rmc_{\,\nu}}+
(\tilde\ppi{^\rmc_{\,\mu}}-\tilde\ppi{^\rmf_{\,\mu}})
\left(\epsilon\,\delta^\mu_\nu+A_\rmc\frac{\partial\epsilon}
{\partial A_\rmc}\uu^\mu\tt_\nu\right)\, ,\fe}
while for the coefficient in the final term of (\ref{46}) we shall 
also obtain a non-trivial adjustment given by
{\be \LmP{^\nu_{\,\rm _A}}=\tilde\LmP{^\nu_{\,\rm _A}}
-(\tilde\ppi{^\rmc_{\,\mu}}-\tilde\ppi{^\rmf_{\,\mu}})\nn_\rmc^{\,\mu}
A_\rmc\frac{\partial\epsilon}{\partial A_\rmc}
\gamm_{\rm_{AB}} \qq^{\rm _B}_{\, ,\rho} 
\gamma^{\rho\nu}\, .\fe}
It can however be seen that the extra terms proportional to
$\partial\epsilon/\partial A_\rmc$ will cancel out
in the final expression (\ref{46a}) for the canonical stress 
energy tensor, whose chemical gauge invariance property
(\ref{46}) is thus confirmed.

\subsection{Vortex pinning and chemical equilibrium}
\label{pineq}

The chemical gauge invariance of the ``free'' (though not the ``confined'')
4-momentum covector is essential for the modelisation of superfluidity, whose 
effect on a mesoscopic scale (large compared with the microscopic lattice 
spacing but small compared with intervortex separation) will be expressed 
by the condition that this covector should have the form of a gradient, 
$\ppi^{\rmf}_{\ \mu}=\hhbar\nabla_{\!\mu}\varpphi$ of a periodic quantum 
phase angle $\varpphi$, with the corollary that the corresponding chemically 
invariant  (and even Milne invariant~\cite{CC03}) ``free'' vorticity tensor 
$\varppi^{\rmf}_{\,\mu\nu}$ will also vanish at the mesoscopic level. At the 
macroscopic level, on a scale large compared with the separation between 
vortices, this vorticity will have a large scale average value that will not 
vanish, but to be compatible with  a mesoscopic fibration by 2 dimensional 
vortex lines it must still be algebraically restricted by the degeneracy 
condition
{\be \varppi^{\rm f}_{\,\mu[\nu}\varppi^{\rm f}_{\,\rho\sigma]}=0
\, .\label{(36)}\fe}

This condition will automatically be satisfied by the field equations that 
will result from the relevant variational principle, which will require that 
the vanishing perturbation condition (\ref{(5)}) should be satisfied, not 
for independent variations of all four of the displacement fields 
$\xi_{\rm f}^{\,\mu}$,  $\xi_{\rm c}^{\,\mu}$, $\xi_\emptyset^{\,\mu}$, 
$\xi^\mu$, which in view of the constraint (\ref{(31)}) would lead to 
overdetermination, but just for variations satisfying the corresponding 
restraint (\ref{11b}). This requirement entails the vanishing, not of all 
four of the relevant force densities $\ff^{\rmf}_{\,\mu}$, 
$\ff^{\rmc}_{\,\mu}$, $\ff^\emptyset_{\,\mu}$ ,  $\ff^{\rms}_{\,\mu}$ 
given by the prescriptions (\ref{(22)}) and (\ref{(25)}), but only of the 
free force density $\ff^{\rmf}_{\,\mu}$  and of the almalgamated force 
density $ \ff^{\rmI}_{\,\mu}$ given by (\ref{12}) that acts on the part that 
is convected with the ionic lattice. 

According to the general formula (\ref{(22)}), the vanishing of the free 
force density $\ff^\rmf_{\,\mu}$ entails not only the separate conservation
condition
{\be \nabla_{\!\nu}\nn_{\rm f}^{\, \nu}=0\, ,\label{(41)}\fe}
for the free part $\nn_\rmf^{\,\mu}$ of the current density but also a 
dynamic equation of the familiar form
{\be \varppi^{\rm f}_{\,\mu\nu}\nn_{\rm f}^{\,\nu}=0\, ,\label{(38)}\fe}
which using  (\ref{(22a)})can be seen to imply  
\be
\nn_{\rm f} \Libra \varppi^{\rm f} = 0 \,\, , 
\fe
which is interpretable as meaning that the vortex lines are ``frozen'' into
the free fluid in the sense of being dragged along with it. It follows,
according to (\ref{(36)}) that there will be no dissipation in the sense 
that the entropy will be conserved,
{\be \nabla_{\!\nu}\ses^{\nu} =0\, .\label{(37)}\fe}

Instead of postulating the strict application of the variation principle,
we can obtain a non-dissipative model of another physically useful kind
by postulating that the vortices are ``pinned'' in the sense of being frozen, 
not into the fluid but into the atomic lattice structure, which means 
replacing (\ref{(38)}) by a dynamical equation of the analogous form 
 {\be \varppi^{\rm f}_{\,\mu\nu}u^{\nu}=0\, ,\label{(39)}\fe}
which unlike (\ref{(38)}) is unaffected by the chemical
base transformations considered in Subsection \ref{chemadj}.

Whichever of the alternative possibilities (\ref{(38)}) or (\ref{(39)}) is 
adopted, the degeneracy condition (\ref{(36)}) will evidently be satisfied,
 and more particularly the left hand side of the identity (\ref{(33)}) will 
vanish, which means that in order to satisfy the entropy conservation
condition (\ref{(37)})  the model must be such as to satisfy the condition
{\be ({\calE}^{\rm c} - {\calE}^{\rm f}) 
\nabla_{\!\nu}\nn_{\rm f}^{\, \nu}=0\, .\label{(40)}\fe}
In the strictly variational case this requirement  will obviously  be 
implemented by the separate conservation condition (\ref{(41)}), which 
will be physically realistic in the context of high frequency oscillations. 
However in applications to slow long term variations it will be more 
realistic to use a model based on the alternative possibility, namely 
that of chemical equilibrium in the solid's rest frame, as given by the 
relation
{\be {\calE}^{\rm c} = {\calE}^{\rm f}\, ,
\label{(42)}\fe}
which, like (\ref{(39)}) can be seen to be unaffected by the chemical
base transformations considered in Subsection \ref{chemadj}.

Thus, depending on how we choose between the alternatives (\ref{(38)}) or 
(\ref{(39)}), and between the alternatives (\ref{(41)}) or (\ref{(42)}), we 
can use the same Lagrangian master function $\Lambda$ for the specification 
of 4 different kinds of non-dissipative model, which are categorisable as 
unpinned with separate conservation or chemical equilibrium, and pinned with 
separate conservation or chemical equilibrium.

\section{Specification of the action}
\label{EqState}
\subsection{External action contribution}

To apply the procedure described above, one needs to prescribe a
(primary) equation of state specifying the functional dependence
of the Lagrangian  master function $\LLambda$ on the relevant 
independent variables.

In accordance with the general principles described in the preceding
work~\cite{CCC}, the relevant Lagrangian will be decomposible in the form
{\be \LLambda=\LLambda_{\rm_{ext}}+\LLambda_{\rm_{int}} \, ,\fe}
in which the  gauge dependent external part is given in terms of a 
fixed baryon mass parameter  $\mm$ by an expression of the familiar 
form
{\be \LLambda_{\rm_{ext}}=\frac{1}{2} \mm\left(\nn_{\rm c}\,\vv_{\rmc}^{\,2}
+\nn_{\rm f}\,\vv_{\rm f}^{\,2}\right)
-\mm \nn_{\rm b}\,\pphi\, .\label{50}\fe}
The confined particle 3-velocity $\vv_\rmc^{\,\nu}$ and the superfluid 
3-velocity $\vv_\rmf^{\,\nu}$ are defined here in terms of the 
corresponding 4-velocities  $\uu_\rmc^{\,\nu}$ and  $\uu_\rmf^{\,\nu}$ 
in the usual way, by setting
{\be \nc^{\,\nu}=\nc\uu_\rmc^{\,\nu}=\nc(\eg^\nu+\vv_\rmc^{\,\nu})
\, ,\hskip 1 cm \nf^{\,\nu}=\nf\uu_{\rm f}^{\,\nu}=
\nf(\eg^\nu+\vv_{\rm f}^{\,\nu})\, ,\label{51}\fe}
and the squares in (\ref{50}) are defined by
{\be \vv_{\rmc}^{\,2}=\eeta_{\mu\nu}\vv_{\rmc}^{\,\mu}\vv_{\rmc}^{\,\nu}
\, ,\hskip 1 cm \vv_{\rmf}^{\,2}=\eeta_{\mu\nu}\vv_{\rmf}^{\,\mu}
\vv_{\rmf}^{\,\nu}\, ,\label{50a}\fe}
where $\eeta_{\mu\nu}$ is the uniform rank-3 space metric defined for the 
Galilean frame characterised by the ether frame vector $\eg^\nu$ according 
to the specifications (\ref{5b}).
This means that the kinetic action contribution is deemed  here to be
independent a priori of the material coordinates $\qq^{\rm_A}$ and 
their gradients, to which it will however be related ``on shell'' by the
application a posteriori of the constraint (\ref{(31)}) to the effect 
that the crust frame should coincide with that of the confined particles, 
i.e.
{\be \uu^\nu=\uu_\rmc^{\, \nu}\, ,\hskip 1 cm\
 \vv^\nu=\vv_\rmc^{\, \nu}\, .\label{51a}\fe}
This approach differs from the treatment used for the non-conducting solid 
limit case \cite{CCC}, in which the constraint (\ref{51a}) was imposed 
in advance, but it leads to results that are entirely equivalent on shell. 
The strategy used here is designed so as to satisfy the condition
(\ref{(34a)}) to the effect that
{\be \tt_\mu \LP^\mu_{\ \nu} =0 \, ,\label{34b}\fe}
which has the advantage of ensuring that (as shown in Subsection 
\ref{relac}) it will be fully consistent with the formalism that has
been developed \cite{CC03,CC04,CC05} for the multiconstituent fluid 
limit case. 

In the formulation used here, the external action provides no 
contribution at all to the extra stress term $\LP^\mu_{\ \nu}$ so --
as in the multiconstituent fluid case \cite{CC04} -- the external 
contribution to the stress energy tensor will be given by the simple 
formula
{\be \TT^{\,\mu}_{\!_{\rm ext}\nu}=\nf^{\,\mu}\pp^\rmf_{\, \nu}
+\nc^{\,\mu}\pp^\rmc_{\, \nu}-\mm\,\pphi\,\nn_{\rm b}^{\,\mu}\tt_\nu
= \nf^{\,\mu}\ppi^{\,\rmf}_{\!_{\rm ext}\nu}+\nc^{\,\mu}
\ppi^{\,\rmc}_{\!_{\rm ext}\nu}\, ,\label{86}\fe}
in which free and confined  kinematic momentum covectors are
given respectively by
{\be \pp^\rmf_{\, \nu}=\mm (\eeta_{\nu\mu}\vv_{\rmf}^{\,\mu}-\frac
{_1}{^2}\vv_\rmf^{\,2}\tt_\nu)\, ,\hskip 1 cm \pp^\rmc_{\, \nu}=
\mm (\eeta_{\nu\mu}\vv_{\rmc}^{\,\mu}-\frac
{_1}{^2}\vv_\rmc^{\,2}\tt_\nu)\, ,\label{86a}\fe}
and the corresponding non-local external momentum covectors are
given by
{\be\ppi^{\,\rmf}_{\!_{\rm ext}\nu}=\pp^\rmf_{\, \nu}-\mm\,\pphi\,
\tt_\nu\, ,\hskip 1 cm\ppi^{\,\rmc}_{\!_{\rm ext}\nu}=\pp^\rmc_{\, \nu}
-\mm\,\pphi\,\tt_\nu\, .\label{86b}\fe}

Since, according to (\ref{20b}) the internal part will not contribute, 
the 3-momentum density (\ref{20a}) will consist just of the external
part, which evidently provides the (chemicaly invariant)
result
{\be \PPi^\nu=\mm (\nf \vv_\rmf^{\,\nu}+\nc\vv_\rmc^{\,\nu})
=\mm\, \eeta^\nu_{\,\mu}\nn_{\rm b}^{\, \mu}\, .\fe}

\subsection{Internal action contribution}

It is to be remarked that the Galileian frame independent difference 
$\vv_{\rm f}^{\,\nu}-\vv^{\nu}$  determines a corresponding --
purely spacelike -- relative current vector,
{\be \nn_{^\perp}^{\,\mu}=\nn_\rmf (\vv_{\rm f}^{\,\nu}-\vv^{\nu})
=\nf^{\,\nu}-\nf\uu^\nu\, ,\hskip 1 cm 
\nn_{^\perp}^{\,\mu} \tt_\mu=0\, ,\label{52}\fe}
which as well as being unaffected by changes of the Galilean frame 
 is also unaffected by chemical base transformations of the form 
(\ref{44}), which will simply give $\tilde\nn_{^\perp}{^\nu}=\nn_{^\perp}^\nu$.

The internal contribution $\LLambda_{\rm_{int}}$ in (77) has to
be independent of the choice of the Galilean ether frame  vector $\eg^\mu$. 
This means that at a given material base location, as specified by the 
fields $\qq^{\rm_A}$, this internal contribution will depend only on the 
scalars $ \ses=\ses^\nu \tt_\nu ,$ $\nn_{\rm f}=\nn_{\rm f}^\nu \tt_\nu$ and 
$\nn_{\rm c}=\nn_{\rm c}^\nu \tt_\nu$ and on the material projections 
{\be \nn^{\rm _A}=\nn_{^\perp}^{\,\mu}
\qq^{\rm_A}_{\, ,\mu}=\nn_{\rm f}^{\,\mu}
\qq^{\rm_A}_{\, ,\mu}\, ,\hskip 1 cm
\gamm^{\rm_{AB}}=\ggamma^{\mu\nu}
\qq^{\rm_A}_{\, ,\mu}\qq^{\rm_B}_{\, ,\nu}\, ,\label{53}\fe}
so that its generic variation will be given by
{\be \delta\LLambda_{\rm_{int}}=\frac{\partial\LLambda_{\rm_{int}}}
{\partial \ses}\delta \ses+\frac{\partial\LLambda_{\rm_{int}}}
{\partial \nn_{\rm c}}\delta  \nn_{\rm c}+\frac{\partial
\LLambda_{\rm_{int}}}{\partial \nn_{\rm f}}\delta  \nn_{\rm f}+
\frac{\partial\LLambda_{\rm_{int}}}{\partial \nn^{\rm_A}}\delta  
\nn^{\rm_A}+\frac{\partial\LLambda_{\rm_{int}}}{\partial 
\gamm^{\rm_{AB}}}\delta  \gamm^{\rm_{AB}}+\frac{\partial
\LLambda_{\rm_{int}}}{\partial \qq^{\rm_A}}\delta \qq^{\rm_A}
\, .\label{58}\fe}

This provides an associated ``convective'' variation \cite{C80} (in 
which, with respect to appropriately dragged coordinates, both 
$\qq^{\rm_A}$ and $\delta \qq^{\rm_A}_{\,\nu}$ are held constant) 
of the form
{\be \delta_{\!\rm _c}\LLambda_{\rm_{int}}=\frac{\partial
\LLambda_{\rm_{int}}}{\partial \ses}\delta_{\!\rm _c} \ses
+\frac{\partial\LLambda_{\rm_{int}}}{\partial \nn_{\rmc}}\delta_{\!\rm _c} 
\nn_{\rmc}+\frac{\partial\LLambda_{\rm_{int}}}{\partial \nn_{\rm f}}
\delta_{\!\rm _c} \nn_{\rm f}+\frac{\partial\LLambda_{\rm_{int}}}{\partial 
\nn_{^\perp}^{\,\nu}}\delta_{\!\rm _c}  \nn_{^\perp}^{\,\nu}+\frac
{\partial\LLambda_{\rm_{int}}}{\partial \ggamma^{\mu\nu}}\delta_{\!\rm _c}  
\ggamma^{\mu\nu}\, ,\label{80}\fe}
in terms of tensorial coefficients defined by 
{\be \frac{\partial\LLambda_{\rm_{int}}}{\partial \nn_{^\perp}^{\,\nu}}
=\frac{\partial\LLambda_{\rm_{int}}}{\partial \nn^{\rm_A}}\qq^{\rm_A}_{,\nu}
\, ,\hskip 1 cm\frac{\partial\LLambda_{\rm_{int}}}{\partial\ggamma^{\mu\nu}}
=\frac{\partial\LLambda_{\rm_{int}}}{\partial 
\gamm^{\rm_{AB}}}\qq^{\rm_A}_{,\nu}\qq^{\rm_B}_{,\mu}
\, .\label{81}\fe}

For the purpose of the present analysis, what we need is the
corresponding ``live'' variation, as carried out at a fixed position 
in a fixed background, for which we obtain
{\be \delta^{_\heartsuit\!}\LLambda_{\rm_{int}}\!=\!\frac{\partial
\LLambda_{\rm_{int}}}{\partial \ses}\delta\ses\!+\!\frac{\partial
\LLambda_{\rm_{int}}}{\partial \nn_{\rm c}}\delta\nc\!+\! 
\frac{\partial\LLambda_{\rm_{int}}}{\partial \nn_{\rm f}}\delta
\nf\!+\!\frac{\partial\LLambda_{\rm_{int}}}{\partial \nn^{\rm_A}}
\delta\left(\qq^{\rm_A}_{,\nu}\nf^\nu\!
\right)\!+\!2\frac{\partial\LLambda_{\rm_{int}}}{\partial 
\gamm^{\rm_{AB}}}\ggamma^{\mu\nu}\qq^{\rm_B}_{,\mu}
\delta\qq^{\rm_A}_{,\nu}\!+\!\frac{\partial
\LLambda_{\rm_{int}}}{\partial \qq^{\rm_A}}\delta \qq^{\rm_A}
 .\label{82}\fe}
We now use the constraint $\uu^\nu\delta\qq^{\rm_A}_{,\nu}
=-\qq^{\rm_A}_{,\nu}\delta\uu^\nu$ to recombine the terms in such a 
way as to obtain a coefficient $\LmP^{\, \mu}_{\,\rm_{A}}$
satisfying the condition (\ref{(34a)}) in an expression of the standard
form
{\be \delta^{_\heartsuit\!}\LLambda_{\rm_{int}}=-\Thet\,\delta\ses+
\cchi^\rmc_{\,\nu}\delta\nc^\nu+\cchi^\rmf_{\,\nu}\delta\nf^\nu
+\LmP^{\, \nu}_{\,\rm_{A}}\delta\qq^{\rm_A}_{,\nu}+\frac{\partial
\LLambda_{\rm_{int}}}{\partial \qq^{\rm_A}}\delta \qq^{\rm_A} 
\, ,\label{83}\fe}
It can be seen that the required (purely spacelike)
value of  $\LmP^{\, \mu}_{\,\rm_{A}}$ will be given by
{\be\LmP^{\, \mu}_{\,\rm_{A}}=\frac{\partial\LLambda_{\rm_{int}}}
{\partial \nn^{\rm_A}}\nn_{^\perp}^{\,\mu}+2\frac{\partial
\LLambda_{\rm_{int}}}{\partial \gamm^{\rm_{AB}}}\qq^{\rm_B}_{,\nu}
\ggamma^{\mu\nu} \, ,\label{84}\fe}
while the corresponding expressions for the internal contributions
to the 4-momentum covectors will be given by
{\be \Thet=-\frac{\partial\LLambda_{\rm_{int}}}
{\partial \ses}\, ,\hskip 1 cm
\cchi^\rmc_{\,\nu}=\frac{\partial\LLambda_{\rm_{int}}}
{\partial \nn_{\rmc}}\tt_\nu-\frac{\nf}{\nc} \frac
{\partial\LLambda_{\rm_{int}}}{\partial \nn_{^\perp}^{\,\nu}}
\, ,\hskip 1 cm \cchi^\rmf_{\,\nu}=\frac{\partial\LLambda_{\rm_{int}}}
{\partial \nn_{\rmf}}\tt_\nu+ \frac{\partial\LLambda_{\rm_{int}}}
{\partial \nn_{^\perp}^{\,\nu}}\, ,\label{85}\fe}
which implies that we shall have
{\be \frac{\partial\LLambda_{\rm_{int}}}
{\partial \nn_{\rmc}}=\uu^\nu\cchi^\rmc_{\,\nu}\, ,\hskip 1 cm
\frac{\partial\LLambda_{\rm_{int}}}{\partial \nn_{\rmf}}
=\uu^\nu\cchi^\rmf_{\,\nu}\, .\label{8-}\fe}

It follows, according to the canonical formula (\ref{(21)})
that the internal contribution to the stess energy tensor will
be given by
{\be \TT^{\,\mu}_{\!_{\rm int}\nu}=-\Thet\,\ses^\mu\tt_\nu
+\cchi^{\rmc}_{\  \nu}\nc^{\,\mu}+\cchi^{\rmf}_{\  \nu}\nf^{\,\mu}
+\PPsi\,\delta^\mu_{\ \nu}-\LP^\mu_{\ \nu}\, ,\label{87}\fe}
with
{\be\LP^{\mu}_{\ \nu}=\frac{\partial\LLambda_{\rm_{int}}}{\partial 
\nn_{^\perp}^{\,\nu}}\nn_{^\perp}^{\,\mu}+2\frac{\partial
\LLambda_{\rm_{int}}}{\partial \ggamma^{\nu\rho}}\ggamma^{\rho\mu} 
\, ,\label{87a}\fe}
and with the generalised presure scalar $\PPsi$ given, as in the 
multiconstituent fluid case, by
{\be \PPsi=\LLambda_{\rm_{int}}+\Thet\,\ses-\cchi^{\rmf}_{\  \nu}
\nf^{\,\nu}-\cchi^{\rmc}_{\  \nu}\nc^{\,\nu}\, .\label{88}\fe}

The identity (\ref{20c}) ensures the symmetric and strictly spacelike 
nature of the ensuing  pressure tensor, as given in accordance with 
(\ref{20e}) by
{\be \PP_{\!_{\rm int}}^{\,\mu\nu}= \TT^{\ \mu}_{\!_{\rm int}\sigma}
\ggamma^{\sigma\nu}\, ,
\label{89}\fe}
which must also be symmetric. It will be obtainable according to the
prescription
{\be  \PP_{\!_{\rm int}}^{\,\mu\nu}=\left(\ses\Thet-
(\nf\cchi^\rmf_{\,\sigma}+\nc\cchi^\rmc_{\,\sigma})\uu^\sigma
- \cchi^\rmf_{\,\sigma}\nn_{^\perp}^{\,\sigma}\right)
\ggamma^{\mu\nu}-\SeS^{\mu\nu}\, ,\hskip 1 cm \SeS^{\mu\nu}=
\ggamma^{\mu\rho}\ggamma^{\nu\sigma}\SeS_{\rho\sigma}\, ,\label{89b}\fe}
in which the stress contribution $\SeS_{\mu\nu}$ is given by an 
expression of the same form as in the non conducting solid case 
\cite{CCC}, namely
{\be \SeS_{\mu\nu}= 2\,\frac{\partial\LLambda_{\rm_{int}}}
{\partial\gamm^{\mu\nu}}-\LLambda_{\rm_{int}}\gamm_{\mu\nu}
=2\,\nn_\rmI\,\frac{\partial(\LLambda_{\rm_{int}}/\nn_\rmI)}
{\partial\gamm^{\mu\nu}}=2 \,\nn_\rmI\,\frac{\partial
(\LLambda_{\rm_{int}}/\nn_\rmI)}{\partial\gamm^{\rm_{AB}}}
\qq^{\rm_A}_{\,,\mu}\qq^{\rm_B}_{\,,\nu}\, .\label{89c}\fe}
This space projected part can be combined with the associated
time projected part, namely the comoving energy current defined by
 {\be \UU^{\rmc\,\mu}_{\!\rm_{int}}=-\TT^{\ \mu}_{\!_{\rm int}\nu}
\uu^\nu\, ,\label{89d}\fe}
to give back the entire internal stress-energy tensor in the form
{\be \TT^{\ \mu}_{\!_{\rm int}\nu}= \PP_{\!_{\rm int}}^{\,\mu\rho}
\gamm_{\rho\nu}-\UU^{\rmc\,\mu}_{\!\rm_{int}}
\tt_\nu    \, ,\label{89e}\fe}
in which $\gamm_{\mu\nu}$ is the time dependent covariant metric that
is given  by
{\be \gamm_{\mu\nu}=\gamm_{\rm_{AB}}\qq^{\rm_A}_{\,,\mu}
\qq^{\rm_B}_{\, ,\nu} \, ,\label{5f}\fe}   where 
the covariant base space metric $\gamm_{\rm_{AB}}$ is such
that 
\be
\gamm_{\rm_{AB}}\gamm^{\rm_{BC}}=\delta^{\rm_{A}}_{\rm_{C}} .
\fe
Using (\ref{(27)}) it can be seen that $\gamm_{\mu\nu}$ is rank 3 : 
\be
\ggamma_{\mu\nu}\uu^\nu=0 .
\fe
One can thus interpret this covariant metric as the rank 3 Euclidian 
metric that is obtained by substituing the ether frame vector defined 
in (\ref{5b}) by the solid's reference frame $\uu^\mu$. As a consequence 
$\gamm_{\mu\nu}$ will be such that
\be
\ggamma_{\mu\rho}\ggamma^{\nu\rho}=\ddelta^\rho_\mu-\uu^\rho\tt_\mu
=\ggamma^\rho_{\,\mu}
\, .\label{5e}\fe

The comoving energy current  $ \UU^{\rmc\,\mu}_{\!\rm_{int}}$ 
(which would be the same as the ordinary internal energy current 
$\UU^{\,\mu}_{\!\rm_{int}}$ in a locally comoving frame -- meaning 
a Galilean gauge with $\eg^\mu=\uu^\mu$ at the position under 
consideration) will be expressible in the form
{\be \UU^{\rmc\,\mu}_{\!\rm_{int}}=\uu^\mu\, \UU_{\!\rm_{int}}
-\frac{\partial\LLambda_{\rm_{int}}}{\partial \nn_\rmf}\,
\nn_{^\perp}^{\,\mu}\, ,\label{89f}\fe}
in terms of the ordinary internal energy $\UU_{\!_{\rm int}}
=-\tt_\mu\TT^{\, \mu}_{\!_{\rm int}\nu} \eg^\nu$ which can be 
seen to be given simply by
{\be \UU_{\!\rm_{int}}=\frac{\partial\LLambda_{\rm_{int}}}
{\partial \nn^{\rm_A}}\,\nn^{\rm_A}-\LLambda_{\rm_{int}}
\, .\label{90g}\fe}
Note that the first term on the right of this formula is not 
something that depends on the elastic solidity, but something that 
-- except in the absence of the entrainment effect \cite{CCH05} 
discussed in Subsection \ref{entrain} -- will  still be present 
even in the fluid limit. This means that (contrary to what has been 
suggested in the literature \cite{Prix02}) in the presence of 
entrainment it will  never be permissible to simply identify the 
internal action density with the negative of the internal energy 
density.

As in the fluid case, we can go on to construct locally defined
material 4-momenta
{\be \mmu^\rmf_{\, \nu} =\pp^\rmf_{\, \nu}+\cchi^\rmf_{\, \nu}
\, , \hskip 1 cm 
\mmu^\rmc_{\, \nu}  =\pp^\rmc_{\, \nu}+\cchi^\rmc_{\, \nu}
\label{90}\fe}
from which, after allowance for the non local effect of gravity,
we finally  get the total momenta for the free and confined particles, 
{\be \ppi^\rmf_{\, \nu} = \mmu^\rmf_{\, \nu} -\mm\,\pphi\,\tt_\nu
\, , \hskip 1 cm 
\ppi^\rmc_{\, \nu} = \mmu^\rmc_{\, \nu}  -\mm\,\pphi\,\tt_\nu
\, ,\label{91}\fe}
in terms of which the complete stress energy tensor will be expressible
as
{\be \TT^{\mu}_{\ \nu}=-\Thet\,\ses^\mu\tt_\nu+\ppi^{\rmf}_{\  \nu}
\nf^{\,\mu}+\ppi^{\rmc}_{\  \nu}\nc^{\,\mu}+\PPsi\delta^\mu_{\ \nu}
-\LP^\mu_{\ \nu}\, .\label{92}\fe}
with
{\be \LP^\mu_{\ \nu} = \SeS^\mu_{\ \nu}+\LLambda_{\rm_{int}} 
\gamm^\mu_{\nu}+\nn_{^\perp}^{\,\mu}\frac{\partial\LLambda_{\rm_{int}}}
{\partial \nn_{^\perp}^\nu}\, .\fe}

\subsection{Elastic energy contribution}

As in the multiconstituent fluid case \cite{CCH05} it is useful
to decompose the internal action
function in the form
{\be \LLambda_{\rm_{int}}=\LLambda_{\rm_{ins}}+\LLambda_{\rm_{ent}}
\, ,\label{100} \fe}
so as to obtain a corresponding decomposition
{\be \SeS^\nu_{\ \mu}=\SeS^{\, \nu}_{\!\rm_{ins}\,\mu}+
\SeS^{\, \nu}_{\!\rm_{ent}\,\mu}\, ,\label{100a}\fe}
in which $\LLambda_{\rm_{ent}}$ and $\SeS^{\, \nu}_{\!\rm_{ent}\,\mu}$
are the parts attributable to entrainment, and
$\LLambda_{\rm_{ins}}$ and $\SeS^{\, \nu}_{\!\rm_{ins}\,\mu}$ are
the static internal contributions that remain when the relative current 
contributions $\nn^{\rm_A}$ are set to zero. For this static part (but 
not for the rest) the action density will just be the opposite of the 
elastic energy density $\EU$ as defined by setting
{\be  \UU{\!\rm_{ins}}=\EU\, , \label{101a}\fe}
which can be seen from   (\ref{90g}) to correspond to
{\be \LLambda_{\rm_{ins}}=-\EU \, .\label{101b}\fe}

It will commonly be convenient to further decompose this static
energy contribution in the form
{\be \EU_{\rm_{ins}}=\EU_{_\bigcirc}+\EU_{\rm_{sol}}\, ,\hskip 1 cm
\EU_{_\bigcirc}=-\LLambda_{_\bigcirc}\, ,\label{54} \fe}
in which $\EU_{_\bigcirc}$ is the part that remains in a relaxed 
configuration for which $\EU$ is minimised for given values of the 
independent current components $\nn_{\rmc}$, $\nn_\rmf$, $\nn^{\rm_A}$,
 and of the determinant of the induced metric $\gamm^{\rm_{AB}}$.  
Fixing this determinant is equivalent to fixing the value of the 
conserved number density $\nn_{\rm I}$ that is specified by (\ref{55}). 
We shall use the notation $\check\gamm{^{\rm_{AB}}}$ and 
$\check\gamm_{\rm_{AB}}$ respectively for the corresponding relaxed 
values of $\gamm^{\rm_{AB}}$ and its inverse $\gamm_{\rm_{AB}}$ (as 
defined by $\gamm_{\rm_{AB}} \gamm^{\rm_{BC}}=\delta^{\rm_C}_{\rm_A}$) 
at which, for given values of $\nn_{\rm I}$, $\nn_{\rmc}$, $\nn_\rmf$, 
the maximisation occurs. Thus substitution of $\check\gamm{^{\rm_{AB}}}$ 
for $\gamm^{\rm_{AB}}$ in the solidity term $\EU_{\rm_{sol}}$ or in the 
total $\EU$ or will give a generically reduced value 
{\be \check \EU_{\rm_{sol}}\leq \EU_{\rm_{sol}}\, ,\hskip
1 cm \check \EU\leq \EU \, ,\label{102a}\fe}
but it will have no effect on the relaxed part, or on the ionic
number density $\nn_{\rm I}$,  for which we simply get
{\be \check\EU_{_\bigcirc}=\EU_{_\bigcirc}\, ,\hskip
1 cm \check \nn_{\rm I} =\nn_{\rm I}\, .\label{102b}\fe}

The relaxed contribution will evidently be of ordinary (albeit
non-barotropic) perfect fluid type with a generic variation of 
the form
{\be \delta \EU_{_\bigcirc}=-\delta\LLambda_{_\bigcirc}=
\Thet_{_\bigcirc}\delta\ses+\cchi^\rmf_{_\bigcirc}\delta\nn_\rmf
+\cchi^\rmc_{_\bigcirc}\delta\nn_\rmc+\cchi^\rms_{_\bigcirc}
\delta\nn_\rmI -\lamb^{\,\rms}_{_{\bigcirc\rm A}}
\delta\qq^{\rm _A}\, ,\label{102c}\fe}
(in which the final term allows for the possibility of built in 
inhomogeneity in addition to the stratification due just to the 
variation of the atomic number ratio $A_\rmc$) with 
{\be \delta\nn_\rmI=\frac{_1}{^2}\,\nn_\rmI\,\gamm_{\rm_{AB}}
\delta\gamm^{\rm_{AB}} + \frac{\partial\nn_\rmI}
{\partial \qq^{\rm_A}}\delta\qq_{\rm_A}\, .\label{103a}\fe}
According to (\ref{85}) the relaxed contribution 
to the pressure tensor (\ref{89b}) will be given by
{\be \PP_{\!_\bigcirc}^{\,\rm_{AB}}= (\Thet_{_\bigcirc}\ses+
\cchi^{\,\rmf}_{_\bigcirc}\nn_\rmf+\cchi^{\,\rmc}_{_\bigcirc}
\nn_\rmc)\gamm^{\rm_{AB}}-\SeS^{\,\rm_{AB}}_{_\bigcirc} \, ,
\hskip 1 cm\SeS^{\,\rm_{AB}}_{_\bigcirc}=(\EU_{_\bigcirc}-
\cchi^{\,\rms}_{_\bigcirc}\nn_\rmI) \gamm^{\rm_{AB}}\, ,\label{104}\fe}
so one finally obtains an expression of the familiar isotropic form
{\be \PP_{\!_\bigcirc}^{\,\rm_{AB}}= \PP_{\!_\bigcirc}\gamm^{\rm_{AB}}
\, ,\hskip 1 cm\PP_{\!_\bigcirc}=
\cchi^{\,\rmf}_{_\bigcirc}\nn_\rmf+
\cchi^{\,\rmI}_{_\bigcirc}\nn_\rmI-\EU_{_\bigcirc}\, ,\hskip 1 cm
\cchi^{\,\rmI}_{_\bigcirc}=\cchi^{\,\rms}_{_\bigcirc}+
A_\rmc\cchi^{\,\rmc}_{_\bigcirc}+\frac{\ses}{\nn_\rmI}\Thet_{_\bigcirc}
\, .\label{104a}\fe}

Let us now consider the solidity contribution $\EU_{\rm_{sol}}$
whose job is to allow for the effect of deviations of 
$\gamm_{\rm_{AB}}$ from its relaxed value $\tilde\gamm_{\rm_{AB}}$
(as determined by the scalars $\nn^\rmf$, $\nn_\rmc$, $\nn_\rmI$).
Such deviations can conveniently be accounted for \cite{CQ72} in 
terms of the constant volume shear tensor whose material base space 
representation is specified as
{\be \sigm_{\rm_{AB}}=\frac{_1}{^2}(\gamm_{\rm_{AB}}
-\check\gamm_{\rm_{AB}})\, ,\label{105a}\fe}
which means that the corresponding space time tensor will be 
given by
{\be \sigm_{\mu\nu}= \sigm_{\rm_{AB}}\qq^{\rm_A}_{,\mu}
\qq^{\rm_B}_{,\nu}=\frac{_1}{^2}(\gamm_{\mu\nu}-\check\gamm_{\mu\nu})
\, .\label{105b}\fe}
In most applications to behaviour of a perfectly elastic (rather than 
plastic or other more complicated) kind, it will be sufficient to use 
an ansatz of quasi-Hookean type \cite{CQ72}, meaning one in which the 
solidity contribution has a homogeneously quadratic dependence on the 
deviation (\ref{105a}) in the sense that it will be given by an 
expression of the form
{\be  \EU_{\rm_{sol}}=\frac{_1}{^2}\,\check{\Shear}{^{\rm_{ABCD}}}
\,\sigm_{\rm_{AB}}\sigm_{\rm_{CD}}
\, ,\label{106}\fe}
with
{\be \check{\Shear}{^{\rm_{ABCD}}}=
\check{\mit\Shear}{^{\rm_{(AB)(CD)}}}=\check{\Shear}{^{\rm_{CDAB}}}
\, ,\label{06a}\fe}
in which $\check{\Shear}^{\rm_{ABCD}}$ is the relevant shear 
elasticity tensor, for which the check symbol is used to indicate 
that, for a given value of the material position coordinates 
$\qq^{\rm_A}$, it depends only on the scalars $\ses ,$ $\nn_\rmf , $ 
$\nn_\rmc$ and (via $\nn_\rmI$) on the relaxed metric 
$\check\gamm_{\rm_{AB}}$. The condition that $\gamma_{\rm_{AB}}$ and 
$\check\gamm_{\rm_{AB}}$ must have the same determinant entails that 
on the 3 dimensional material base the symmetric shear tensor 
$\sigm_{\rm_{AB}}$ will have only 5 (instead of 6) independent 
components, and more specifically that to first order it will be trace 
free with respect to either the actual metric $\gamm_{\rm_{AB}}$ or 
the  relaxed $\check\gamm_{\rm_{AB}}$. It is therefore necessary to 
impose a corresponding restriction to completely fix the specification 
of the solidity tensor $\check{\Shear}{^{\rm_{ABCD}}}$, which can most 
conveniently be done~\cite{CQ72} by requiring that it be trace free 
with respect to the relaxed metric
{\be \check{\Shear}{^{\rm_{ABCD}}\check\gamm_{\rm_{CD}}}=0
\, .\label{06b}\fe}

The specification of the solidity (i.e. shear elasticity) tensor is 
not by itself sufficient to complete the specification of the elastic 
system, as it is also necessary to specify the dependence on $\nn_\rmI$ 
of the relaxed inverse metric $\check\gamm^{\rm_{AB}}$. The simplest 
possibility is that of what has been termed a perfect solid \cite{CQ72},
meaning one at which the elastic structure at each material position 
is isotropic with respect to the relaxed metric, which in that case 
can vary only by a conformal factor. This means that it will be given 
in terms of its value $\gamm_0^{\,\rm_{AB}}$ say at some fixed 
reference value $\nn_0$ say of the ionic number density $\nn_\rmI$ by
{\be \check\gamm^{\rm_{AB}} =(\nn_I/\nn_0)^{2/3}\gamm_0^{\,\rm_{AB}}
\, ,\hskip 1 cm\check\gamm_{\rm_{AB}} =(\nn_0/\nn_\rmI)^{2/3}
\gamm_{0\rm_{AB}}\, .\label{07}\fe} 
In the case of a solid stucture that is isotropic (as will typically be
the case on a macroscopic scale after averaging over randomly oriented
mesoscopic crystals) the rigidity tensor in the quasi Hookean ansatz
will simply have to be given in terms of the relevant scalar shear 
modulus $\check\Shear$ by the formula
{\be \check{\Shear}{^{\rm_{ABCD}}}= 2\check\Shear\,(\check\gamm^{\rm_{A(C}}
\check\gamm^{\rm_{D)B}}-\frac{_1}{^3} \check\gamm^{\rm_{AB}}
\check\gamm^{\rm_{CD}})\, ,\label{107a}\fe}
where the scalar $\check\Shear$ is the rigidity modulus which is usually 
denoted by $\mu$ in the litterature, a symbol which is already being used 
in the general formalism used here to design the material momentum components.
 
Equation (\ref{106}) will then give the simple formula
{\be  \LLambda_{\rm_{sol}} = -\EU_{\rm_{sol}}=-\check\Shear\,\sigm^2
\, ,\label{07c}\fe} 
in which the scalar shear magnitude $\sigm$ is defined by the formula
{\be \sigm^2=\check\gamm^{\rm_{AB}}\gamm^{\rm_{CD}}\sigm_{\rm_{BC}}
\sigm_{\rm_{DA}}-\frac{_1}{^3}(\check\gamm^{\rm_{AB}}
\sigm_{\rm_{AB}})^2\, ,\label{07e}\fe}
in which the final term will in pratice be negligible since of 
quartic order, ${\cal O}\{\sigm^4\}$, in the small $\sigm$ limit that 
is relevant, as the trace is already of quadratic order, 
$\check\gamm^{\rm_{AB}} \sigm_{\rm_{AB}}= {\cal O}\{\sigm^2\}$.
Under these condition  the solidity contribution will provide a 
pressure tensor given by the formula
{\be \PP_{\!\rm_{sol}}^{\,\rm_{AB}}= -\check{\Shear}{^{\rm_{ABCD}}}
\sigm_{\rm_{AB}} +\PP_{\!\rm_{sol}}\gamm^{\rm_{AB}}\, ,\label{08} \fe}
of which the final term is a pressure contribution given by
{\be \PP_{\!\rm_{sol}}=\left(\ses\frac{\partial\Shear}
{\partial\ses}+\nn_\rmf\frac{\partial\Shear}{\partial\nn_\rmf}
+\nn_\rmc\frac{\partial\Shear}{\partial\nn_\rmc}
+\nn_\rmI\frac{\partial\Shear}{\partial\nn_\rmI}+\frac{\Shear}{3}
\right)\sigm^2\, ,\label{08a}\fe}
which, since it is of quadratic order in $\sigm$, will be
negligible compared with the first (linear order) term for most 
practical purposes. (It is to be noted that in the final term of 
(\ref{08a}) the sign given here corrects an error in the sign of 
the corresponding term in the relevant equation (6.19) as written
in the original treatment \cite{CQ72} of the perfect solid model.)
The corresponding pressure adjustment contribution for the canonical 
formula (\ref{87}) will be given by
{\be \LP^\mu_{\ \nu} =\gamm_{\nu\lambda}
\tilde\Shear{^{\lambda\mu\rho\sigma}}\sigm_{\rho\sigma}-\left(
\nn_\rmI\frac{\partial\Shear}{\partial\nn_\rmI}+\frac{4\Shear}{3}
\right)\sigm^2\gamm^\mu_{\ \nu}\, ,\label{08c}\fe}
in which, again, the quadratic order term at the end will be
negligible in practice.

\subsection{Entrainment contribution}
\label{entrain}

In general the entrainment action function $\LLambda_{\rm_{ent}}$ will 
depend, for a given values of $\qq^{\rm_A}$, on the relative current
components $\nn^{\rm_A}=\qq^{\rm_A}_{,\nu}\nn_{^\perp}^{\,\nu}$
as well as on the scalar magnitudes 
$\nn_\rmf$ , $\nn_\rmc$, and the induced metric components
$\gamm^{\rm_{AB}}$ , so its generic variation will have the form
{\be \delta\LLambda_{\rm_{ent}}=-\Thet_{\!\rm_{ent}}\delta\ses
-\cchi^{\,\rmf}_{\!\rm_{ent}}\delta\nn_\rmf- 
\cchi^{\,\rmc}_{\!\rm_{ent}}\delta\nn_\rmc+
\frac{\partial\LLambda_{\rm_{ent}}}{\partial \gamma^{\rm_{AB}}}
\delta\gamma^{\rm_{AB}}+\frac{\partial\LLambda_{\rm_{ent}}}
{\partial \nn^{\rm_A}}\delta\nn^{\rm_A}+\frac
{\partial\LLambda_{\rm_{ent}}}{\partial \qq^{\rm_A}}\delta\qq^{\rm_A}
\, .\label{64}\fe}

The entrainment action $\LLambda_{\rm_{ent}}$ is characterised by the 
condition that it vanishes when the relative current components 
$\nn^{\rm_A}$ are set to zero, so when these components are sufficiently 
small, as will typically be the case, it will be a good approximation to 
take this contribution to have the homogeneous quadratic form
{\be\LLambda_{\rm_{ent}}=\frac{1}{2\nn_\rmf}\mm^{^\perp}_{\rm_{AB}}
\nn^{\rm_A}\nn^{\rm_B}\, ,\label{64c}\fe}
in which the entrainment mass tensor has components
$\mm^{^\perp}_{\rm_{AB}}$, that (like the static action contribution) 
are independent of the current components $\nn^{\rm_A}$, so that the 
corresponding partial derivative in (\ref{64}) will be given by
{\be \frac{\partial\LLambda_{\rm_{int}}}{\partial \nn^{\rm_A}}
=\frac{\partial\LLambda_{\rm_{ent}}}{\partial \nn^{\rm_A}}= 
\frac{1}{\nn_\rmf}\mm^{^\perp}_{\rm_{AB}}\nn^{\rm_B}\, .\label{64d}\fe}

It is conceivable that the relaxed action function might involve a 
built in anisotropy favoring relative currents in some particular 
direction, but in cases of the simplest kind, to which the remainder 
of this subsection and the next will be restricted, this function
$\LLambda_{\rm_{ent}}$ will be of purely fluid type in the sense that 
for given values of $\qq^{\rm_A}$ it will  depend only on the set of 
five scalar magnitudes consisting of $\nn_\rmI ,$ $\nc ,$ $\nf ,$ 
$\ses\ ,$ together with the relative current magnitude $\nn_{^\perp}$ that 
is defined in terms of the (unrelaxed) metric value $\gamma_{\rm_{AB}}$ 
which -- using the material index lowering operation specified by the 
induced metric $\gamm^{\rm_{AB}}$ -- will be given by
{\be \nn_{^\perp}^{\,2} =\nn^{\rm_A}\nn_{\rm_A}\, ,\hskip 1 cm 
\nn_{\rm_A}=\gamm_{\rm_{AB}}\nn^{\rm_B}\, ,\hskip 1 cm \gamm_{\rm_{AB}}
\gamm^{\rm_{BC}}=\delta^{\rm_C}_{\rm_A}\, .\label{110}\fe}
Such a functional dependence provides an expansion
{\be \delta\LLambda_{\rm_{ent}}=-\Thet_{\!\rm_{ent}}\delta\ses
-\cchi^\rmf_{\rm_{ent}}\delta\nn_\rmf-
\cchi^\rmc_{\rm_{ent}}\delta\nn_\rmc-\cchi^\rms_{\rm_{ent}}\delta\nn_\rmI 
+\lamb^{\,\rms}_{\!\rm_{ent\, A}}\delta\qq^{\rm _A}+\frac
{\partial\LLambda_{\rm_{ent}}}{\partial\nn{_{^\perp}^{\,2}}}
\delta\nn{_{^\perp}^{\,2}}\, ,\label{111}\fe}
of similar form to the perfect fluid contribution (\ref{102c}) 
but with an extra term involving a partial derivative that
provides an expression of the form (\ref{64d}) in terms
an isotropic mass tensor given by
{\be \mm^{^\perp}_{\rm_{AB}}=\mm^\rmf_{\,\rmc}\gamm_{\rm_{AB}}\, ,
\hskip 1 cm \mm^\rmf_{\,\rmc}= 2\nn_\rmf\frac 
{\partial\LLambda_{\rm_{ent}}}{\partial\nn{_{^\perp}^{\,2}}}
\, ,\label{112}\fe}
while the other partial derivatives in (\ref{64}) will be given by
{\be \frac{\partial\LLambda_{\rm_{ent}}}{\partial\gamm^{\rm_{AB}}}
=-\frac{_1}{^2}\left(\frac{\mm^\rmf_{\,\rmc}}{\nn_\rmf}\,
\nn_{\rm_A}\nn_{\rm_B} +{\cchi^{\,\rms}_{\!\rm_{ent}}}\,
\nn_\rmI\,\gamma_{\rm_{AB}}\right) \, ,\hskip 1 cm
\frac{\partial\LLambda_{\rm_{ent}}}{\partial \qq^{\rm _A}}
=\lamb^{\rms}_{\!\rm_{ent\,A}}-\cchi^{\,\rms}_{\!\rm_{ent}}
\frac{\partial\nn_\rmI}{\partial\qq^{\rm_A}}\, . \label{113}\fe}
The scalar $\mm^\rmf_{\,\rmc}$ introduced in this way is identifiable 
as the increment $\mm^\rmf_{\,\rmc}=\mm_\star-\mm$ of the effective 
mass $\mm_\star$ of the free baryons (meaning the superfluid neutrons) 
as compared with the ordinary baryonic mass $\mm$. This mass increment 
is expected to be positive (and in some layers large) \cite {CCH05} in 
the solid neutron star crust, but (moderately) negative in the fluid 
layers below.

\subsection{Relaxed action contribution}
\label{relac}

It will be useful for many purposes to replace the decomposition
(\ref{100}) of the internal action density by an alternative
decomposition of the form
{\be\LLambda_{\rm_{int}}=\LLambda_{\rm_{lax}}+\LLambda_{\rm_{sol}}
\, ,\label{120}\fe}
in which the relaxed -- meaning shear independent -- part will 
evidently consist of the combination 
 {\be\LLambda_{\rm_{lax}}=\LLambda_{_\bigcirc}+\LLambda_{\rm_{ent}}
\, .\label{121}\fe}

The use of such a combination is particularly convenient whenever the 
entrainment contribution is of the isotropic type characterised by 
the variation expansion (\ref{111}), in which case  the complete 
relaxed action density will have a variation given by an expansion of 
the analogous form
{\be \delta\LLambda_{\rm_{lax}}=-\Thet_{\!\rm_{lax}}\delta\ses
-\cchi^\rmf_{\rm_{lax}}\delta\nn_\rmf-
\cchi^\rmc_{\rm_{lax}}\delta\nn_\rmc-\cchi^\rms_{\rm_{ent}}\delta\nn_\rmI 
+\lamb^{\,\rms}_{\rm_{lax\, A}}\delta\qq^{\rm _A}
+\frac{\mm^\rmf_{\ \rmc}}{2\nn^\rmf}\,
\delta\nn{_{^\perp}^{\,2}}\, ,\label{122}\fe}
with
$$ \Thet_{\rm_{lax}}=\Thet_{_\bigcirc}\!+\!\Thet^\rmf_{\rm_{ent}}\, , \ \ 
\ \cchi^\rmf_{\rm_{lax}}=\cchi^\rmf_{_\bigcirc}\!+\!\cchi^\rmf_{\rm_{ent}}
\, ,\ \ \  \cchi^\rmc_{\rm_{lax}}=\cchi^\rmc_{_\bigcirc}\!
+\!\cchi^\rmc_{\rm_{ent}}\, ,$$
{\be  \cchi^\rms_{\rm_{lax}}=
\cchi^\rms_{_\bigcirc}\!+\!\cchi^\rms_{\rm_{ent}}\, ,\ \ \
\lamb^\rms_{\rm_{lax\, A}}=\lamb^\rms_{_{\bigcirc A}}\!+\!
\lamb^\rms_{\rm_{ent\, A}}\, .\label{123}\fe}

It can be seen that the relaxed contribution to the ``live'' action 
variation (\ref{83}) will simplify to provide an expression of the form
{\be \delta^{_\heartsuit\!}\LLambda_{\rm_{lax}}=-\Thet_{\!\rm_{lax}}
\delta\ses+\cchi^{\,\rmf}_{\!_{\rm lax}\nu} \, \delta\nn_{\rm f}^{\,\nu}
+ \cchi^{\,\rmc}_{\!_{\rm lax}\nu}\, \delta\nn_{\rm c}^{\,\nu} 
-\cchi^{\rms}\delta^{_\heartsuit\!}\nn_{\rmI}
+\lamb^{\,\rms}_{\rm_{lax\, A}}
\delta\qq^{\rm _A}\, ,\label{60}\fe}
in which  $\cchi^{\,\rmf}_{\!_{\rm lax}\nu}$ and 
$\cchi^{\,\rmc}_{\!_{\rm lax}\nu}$ can be read out as the
relaxed parts of the internal momenta given by (\ref{85}).
Since we can write $ \delta^{_\heartsuit\!}\nn_{\rm I}=\tt_\mu\,
\delta^{_\heartsuit\!}\nn_{\rm I}^{\, \mu}$, it can be seen that, 
 as the analogue of these free and confined particle 4-momentum 
contributions, we shall also be able to read out a corresponding
ionic stratification 4-momentum contribution  
$\cchi^{\,\rms}_{\!_{\rm lax}\nu}$, so as to obtain a complete set
of relaxed internal momentum covectors that will be given by
{\be \cchi^{\,\rmf}_{\!_{\rm lax}\mu}=-\cchi^{\,\rmf}_{\!_{\rm lax}}
\tt_\mu\!+\frac{\mm^\rmf_{\ \rmc}}{\nn_\rmf}\gamm_{\mu\nu}
\nn_{^\perp}^{\,\nu}\, ,\hskip 0.6 cm \cchi^{\,\rmc}_{\!_{\rm lax}\mu}=
-\cchi^{\,\rmc}_{\!_{\rm lax}}\tt_\mu\!-\frac{\mm^\rmf_{\ \rmc}}
{\nn_\rmc}\gamm_{\mu\nu}\nn_{^\perp}^{\,\nu}\, , \hskip 0.6 cm 
\cchi^{\,\rms}_{\!_{\rm lax}\nu}=-\cchi^{\,\rms}_{\!_{\rm lax}}
\tt_\nu\, .\fe}
In the formula (\ref{84}) for the extra stress, it transpires
that the contributions involving the mass increment 
$\mm^\rmf_{\ \rmc}$  (proportional to $\partial
\LLambda_{\rm_{lax}}/\partial\nn_{^\perp}^{\,2}$) will cancel out, 
leaving the expression
{\be \LmP^{\, \nu}_{\!\rm_{lax\,A}}=-\cchi^{\rms}_{\!_{\rm lax}}
\nn_{\rm I} \,\gamm_{\rm_{AB}}\qq^{\rm _B}_{\, ,\mu} \gamma^{\mu\nu}
\, ,\label{63}\fe}
in which the part due to stratification is all that remains.
It can be seen that the corresponding space time tensor
will be given simply by
{\be \LP^{\, \nu}_{\!\rm_{lax}\,\mu}=\cchi^{\rms}_{\!_{\rm lax}}
\nn_\rmI(\uu^\nu \tt_\mu-\ddelta^\nu_{\,\mu})\, .\label{96}\fe}

In terms of their contributions to the total momenta, which
will simply be given by
{\be \ppi^\emptyset\!_{\!_{\rm lax}\nu}=-\Thet\!_{\!_{\rm lax}} 
\tt_\nu\, ,\hskip 1 cm\ppi^{\rmf}\!_{\!_{\rm lax}\nu}
=\cchi^{\rmf}\!_{\!_{\rm lax}\nu}\, ,\hskip 1 cm 
\ppi^{\rmc}\!_{\!_{\rm lax}\nu}=\cchi^{\rmc}\!_{\!_{\rm lax}\nu}
\, ,\hskip 1 cm\ppi^{\rms}\!_{\!_{\rm lax}\nu}=
\cchi^{\rms}\!_{\!_{\rm lax}\nu}\, ,\fe}
the corresponding elastically relaxed force contributions acting on 
the entropy current and on the free and confined particle currents will 
be given by an ansatz of the standard form (\ref{(22)}) which gives
{\be \ff^\emptyset_{\!_{\rm lax} \nu}= 2\ses^{\mu} 
\nabla_{\![\mu}\ppi^\emptyset\!_{\!_{\rm lax}\nu]}+
\ppi^\emptyset_{\!_{\rm lax} \nu} 
\nabla_{\!\mu}\ses^{\,\mu} \, ,\fe} 
{\be \ff^{\rm f}_{\!_{\rm lax} \nu}= 2\nn_{\rm f}^{\,\mu} 
\nabla_{\![\mu}\ppi^{\rm f}\!_{\!_{\rm lax}\nu]}+
\ppi^{\rm f}_{\!_{\rm lax} \nu} 
\nabla_{\!\mu}\nn_{\rm f}^{\,\mu} \, ,\label{70}\fe} 
and 
{\be \ff^{\rm c}_{\!_{\rm lax} \nu}= 2\nn_{\rm c}^{\,\mu} 
\nabla_{\![\mu}\ppi^{\rm c}_{\!_{\rm lax}\nu]}+
\ppi^{\rm c}_{\!_{\rm lax} \nu} \nabla_{\!\mu}\nn_{\rm c}^{\,\mu} 
\, ,\label{71}\fe}
while for the analogously defined force contribution (\ref{(23)}) due to 
stratification acting on the underlying ionic lattice, it can be seen that
similar reasonning leads to an expression of the slightly different form
{\be \ff^\rms_{\!_{\rm lax} \nu}= 2\nn_{\rm I}^{\,\mu} 
\nabla_{\![\mu}\ppi^{\rms}_{\!_{\rm lax}\nu]}+
\lamb^{\,\rms}_{\!\rm_{lax}\nu}\, ,\hskip 1 cm
\lamb^{\,\rms}_{\!\rm_{lax}\nu}=
\lamb^{\,\rms}_{\!\rm_{lax\,A}}\qq^{\rm_A}_{\, ,\nu}\, .\label{72}\fe}

In terms of the amalgamated ionic 4-momentum  contribution defined by
 {\be \ppi^{\rmI}\!_{\!_{\rm lax}\nu}= \ppi^{\rms}\!_{\!_{\rm lax}\nu} 
+A_\rmc\,\ppi^{\rmc}\!_{\!_{\rm lax}\nu}+({\ses}/{\nn_\rmI})
\ppi^\emptyset\!_{\!_{\rm lax}\nu}\, ,\label{67}\fe}
the  corresponding  contribution 
{\be  \ff^\rmI_{\!_{\rm lax} \nu}=\ff^\rms_{\!_{\rm lax} \nu}+
\ff^\rmc_{\!_{\rm lax} \nu}+\ff^\emptyset_{\!_{\rm lax} \nu}
\, ,\label{73}\fe}
to the amalgamated ionic force density (\ref{12}) can be seen to
be expressible directly  by the formula
{\be \ff^\rmI_{\!_{\rm lax} \nu}= 2\nn_{\rm I}^{\,\mu} 
\nabla_{\![\mu}\ppi^{\rmI}_{\!_{\rm lax}\nu]}+\nn_{\rmI}^{\,\mu}
\ppi^{\rm c}_{\!_{\rm lax}\mu}\nabla_{\!\nu}A_\rmc-\nn_\rmI
\Thet_{\!_{\rm lax}}\!\nabla_{\!\nu}\big({\ses}/{\nn_\rmI}\big)
+\lamb^{\,\rms}_{\!\rm_{lax}\nu}\, , \label{74}\fe}
while the associated contribution to the stress energy tensor 
(\ref{(21)}) will be given neatly by
{\be \TT^{\ \mu}_{\!_{\rm lax} \nu}=(\LLambda_{_{\rm lax}}-
\nn_{\rmf}^{\,\rho}\ppi^{\rmf}_{\!_{\rm lax} \rho}-\nn_{\rmI}^{\,\rho}
\ppi^{\rmI}_{\!_{\rm lax} \rho})\delta^\mu_{\ \nu}+\nn_{\rmf}^{\,\mu} 
\ppi^{\rmf}_{\!_{\rm lax}  \nu}+\nn_{\rmI}^{\,\mu} 
\ppi^{\rmI}_{\!_{\rm lax} \nu}\, .\label{75}\fe}

\subsection{Complete description for perfect conducting solid.}

Replacing the relaxed contribution (\ref{67}) by the corresponding
total ionic 4-momentum covector given (in terms of the confined
atomic number $A_\rmc$)  by
{\be \ppi^{\rmI}_{\,\nu}= \ppi^{\rms}_{\,\nu}+A_\rmc\,
 \ppi^{\rmc}_{\,\nu}-({\ses}/{\nn_\rmI})\Thet\,\tt_\nu
\, , \label{67a}\fe}
we can apply an analogous tidying up operation to the complete
stress energy tensor (\ref{(21)}) which will thereby aquire the form
{\be \TT^{\mu}_{\ \nu}=(\LLambda-
\nn_{\rmf}^{\,\rho}\ppi^{\rmf}_{\,\rho}-\nn_{\rmI}^{\,\rho}
\ppi^{\rmI}_{\, \rho})\delta^\mu_{\ \nu}+\nn_{\rmf}^{\,\mu} 
\ppi^{\rmf}_{\, \nu}+\nn_{\rmI}^{\,\mu} 
\ppi^{\rmI}_{\, \nu}-\LP^{\, \mu}_{\!\rm_{sol}\,\nu}\
\, ,\label{115}\fe}
in which the only manifest allowance for the effects of solid 
rigidity is in the final term. In the case of a perfect 
conducting solid -- meaning one whose structure is fully isotropic
with respect to the relaxed metric -- this final term
$\LP^{\, \mu}_{\!\rm_{sol}\,\nu}$ will be given by the formula 
(\ref{08c}) which will be expressible to first order in the shear 
amplitude $\sigm$ by a prescription of the simple form
{\be \LP^{\, \mu}_{\!\rm_{sol}\,\nu}=2\Shear\,\gamma^{\mu\rho}
\sigm_{\rho\nu}+{\cal O}\{\sigm^2\}\, ,\label{116}\fe}
in which $\Shear$ is the ordinary shear modulus (which elsewhere is 
commonly denoted by the symbol $\mu$ that, in the present context, 
has already been used for the designation of the material momentum 
components). In principle, the other terms in (\ref{115})  will also be 
influenced by the solid rigidity contribution $\LLambda_{\rm_{sol}} ,$ 
but since such effects too will be of quadratic order in $\sigm ,$ 
they will be effectively negligible in applications of the usual 
kind, in which deviations from an elastically relaxed configuration 
are small. This means that for practical purposes, in a conducting
solid of the perfect (meaning intrinsically isotropic) type, the 
deviation from behaviour of (multiconstituent) fluid type \cite{CC04} 
will be entirely contained in the extra term given by (\ref{116}).

\section{Non conservative generalisation}
\label{Dissip}
\subsection{Dissipative interpolation}

In the non-dissipative models described above, it can be seen that the 
introduction of the entropy density $\ses$ as an indpendent variable was 
in practice redundant, since its effects could be allowed for simply by 
a readjustment of the stratification, because the ratio $\ses/\nn_\rmI$ 
was fixed on each material world line so that it depended only on the 
material  base space variables  $\qq^{\rm _A}$

The reason for taking the trouble of introducing it is that the entropy 
density will acquire a non-trivial role as soon as these convectively 
conducting solid models are  generalised to allow for dissipative effects 
in the manner that has recently been described in detail \cite{CCH05} for 
the fluid case. The second law of thermodynamics tells us that, in a
dissipative application, the entropy current $\ses^\nu$ need not satisfy 
the conservation condition (\ref{(37)}), but that -- for a system that 
is closed in the sense of being thermodynamically isolated from the
rest of the universe -- the model must be such as to ensure
satisfaction of the inequality
{\be  \nabla_{\!\nu}\ses^{\nu}\geq 0 \, .\label{(120)}\fe}

On the assumption that it is isolated not just thermodynamically but
in the stronger mechanical sense of (\ref{(28)}), we have seen that a 
system characterised by an action of the kind presented above, and thus 
by a stress energy tensor of the form (\ref{46a}), must automatically 
satisfy the energy conservation identity (\ref{(33)}). Subject to the 
usual presumption that the temperature $\Thet$ is positive, this means 
that the second law requirement (\ref{(120)}) will be expressible for 
such a model as
{\be ({\calE}^{\rm c} - {\calE}^{\rm f})\nabla_{\!\nu}
\nn_\rmf^{\, \nu}+\nn_{\rm f}^{\,\mu}\varppi^{\rm f}_{\,\mu\nu}
\uu^\nu\geq 0  \, .\label{(121)}\fe}
The simplest and most obviously natural way of satisfying this
positivity requirement starts by postulating that the two terms
in (\ref{(121)}) are separately positive. For the first term
this leads to ansatz to the effect that relevant neutron
fluid creation rate should be given  by an expression of the form
{\be \nabla_{\!\nu}\nn_{\rm f}^{\, \nu}=
\kkappa\,({\calE}^{\rm c} - {\calE}^{\rm f}) \, .\label{(122)}\fe}
in terms of some positive transfusion coefficient $\kappa$ that might 
be expected to be a sensitive function of the temperature, $\Thet ,$
on which the weak interactions that would be involved are known 
\cite{Haensel92} to be highly dependent. (It is to be noticed 
that the formula (\ref{(122)}) will be preserved by the chemical 
gauge transformation (\ref{44}) for any fixed value of the 
adjustment parameter $\epsilon$.) 

The treatment of the second term in (\ref{(121)}) is not so simple, 
because it is necessary to respect the degeneracy requirement 
(\ref{(36)}) that ensures that the mesoscopically averaged vorticity 
$\varppi^{\rm f}_{\,\mu\nu}$ is orthogonal to two-surfaces that represent 
the flux of quantised vorticity tubes with generators of the form 
$\uu_{\rm f}^{\, \nu}+\Vv_\rmf^{\,\nu}$ for some spacelike vector 
$\Vv_\rmf^{\,\nu}$ such that
{\be \varppi^{\rm f}_{\,\mu\nu}(\uu_{\rm f}^{\, \nu}+\Vv_\rmf^{\,\nu})
=0\, , \hskip 1 cm \Vv_\rmf^{\,\nu}\tt_\nu=0 \, .\label{(123)}\fe}
In terms of such a vector, the positivity condition on the second
term in (\ref{(121)}) will be expressable (assuming positivity of
$\nn_\rmf$) as
{\be \uu^\mu \varppi^{\rmf}_{\,\mu\nu} \Vv_\rmf^{\,\nu}\geq 0
\, ,\label{(124)}\fe}
so a vector of the required form will be given by an expression of 
the form 
{\be \Vv_\rmf^{\,\nu}=-\frac{\cc_\rmr}{\ww^\rmf}\ggamma^{\nu\rho}
\varppi^{\rm f}_{\,\rho\sigma}\uu^\sigma\, ,\label{(125)}\fe}
in which ${\cc_\rmr}$ is a positive drag coefficient that has been
adjusted so as to be dimensionless by the inclusion of the denominator
$\ww^\rmf$, which is defined as the magnitude of the (spacelike)
vorticity vector $\ww^{\rmf\mu}=\frac{1}{2}\varepsilon^{\mu\nu\rho}
\varppi^{\rm f}_{\,\nu\rho}.$ This provides a superfluid equation of 
motion of the form
{\be\uu_\rmf^{\,\rho}\varppi^{\rm f}_{\,\rho\sigma}\ggamma^{\sigma\mu}
=\cc_\rmr\ww^\rmf\vv_\rmr^{\, \mu}\, ,\label{(126)}\fe}
in which $\vv_\rmr^{\,\mu}$ is a relative flow velocity, namely that
of the medium relative to the vortex sheets, as defined \cite{CC05} in 
terms of the relevant orthogonal projection operator by 
{\be \vv_\rmr^{\, \mu}=\pperp^{\!\mu}_{\, \nu}\uu^\nu\, ,\hskip
1 cm \pperp^{\!\mu}_{\, \nu}=(\ww^\rmf)^{-2}\ggamma^{\mu\rho}
\ggamma^{\sigma\tau}\varppi^{\rm f}_{\,\nu\sigma}
\varppi^{\rm f}_{\,\rho\tau} \, .\label{(127)}\fe}
On the basis of work by Jones \cite{Jones90}, a formula giving a rather 
low value for the required drag coefficient $\cc_\rmr$, in the low 
temperature limit, as a function of the relevant densities, $\nn_\rmI,$ 
$\nn_\rmc, $ $\nn_\rmf, $ has been provided by  Langlois {\it et al.} 
\cite{Lan98}, but it depends on microscopic parameters that are 
difficult to evaluate, and its validity is
in any case a subject of controversy. In the very different picture 
developped by  Alpar {\it et al.} \cite{Alpar89} it has been
suggested that the result will be highly temperature dependent, and
that the effect will be describable as (nonlinear) ``creep'' rather 
than ordinary (linear) drag, in the sense that the appropriate 
coefficient $\cc_\rmr$ will not just depend on the relevant scalar 
densities as well as the temperature $\Thet$, but that it will also be 
strongly dependent on the magnitude $\vv_\rmr$ of the relative velocity 
(\ref{(127)}) (according to a formula of the form $ \cc_\rmr \propto 
\vv_\rmr^{\, -1} {\rm arcsinh}\{\vv_\rmr/\bar \vv_\rmr\}$ for some 
velocity independent -- but temperature sensitive --  quantity 
$\bar \vv_\rmr$).

It is to be remarked that the chemical reaction rate formula 
(\ref{(122)}) provides an interpolation between the non dissipative
limits $\kkappa\rightarrow\infty$, namely the thermal equilibrium
case, and the variational case  $\kkappa\rightarrow 0$, 
for which free and confined particle currents are separately conserved.
Similarly the drag or creep formula (\ref{(126)}) provides an
interpolation between the non dissipative limits $\cc_\rmr
\rightarrow\infty$, namely the perfect pinning case, and 
variational case  $\cc_\rr\rightarrow 0$, for which the vortices are
freely transported by the superfluid flow. 
 
\vfill\eject
\subsection{Conclusion}

Subject to the prescription of two suitable secondary equations of state
for the coefficients $\kkappa$ and $\cc_\rmr$ as functions of the
relevant variables, the foregoing equations (\ref{(122)}) and 
(\ref{(126)}) constitute a complete system of equations of motion for 
the nine independent component variables (which can be considered to be 
the four superfluid current components $\nn^\nu$, and the five materially 
convected components $\qq^{\rm_A}$ $\nn_\rmc$ and $\ses$) when used in 
conjunction with the baryon conservation condition (\ref{(32)}) and the 
condition (\ref{(28)}) of conservation of the stress energy tensor
$\TT^\mu_{\ \nu}$ that is obtained, according to (\ref{115}), from
the the master function $\LLambda$, whose prescription, by a suitable
primary equation of state is described in Section \ref{EqState}.

The stress energy conservation condition (\ref{(28)}) expresses
the requirement that the system should be effectively isolated, not
just mechanically, but also thermally. This last requirement will not be 
entirely realistic when the transfusive adjustments governed by 
(\ref{(122)}) are taking place \cite{Haensel92}, since the beta processes
involved will create neutrinos for which the stellar medium will be 
effectively tranparent so that instead of being locally confined they 
will rapidly escape from the system. The ensuing heat loss can easily be 
formally taken into account \cite{CC05} by replacing (\ref{(28)}) by a 
generalisation in which there is an extra term so, that it takes the form
{\be \nabla_{\!\mu}\TT^\mu_{\ \nu}=-\rrho\nabla_{\!\nu}\pphi+{\calQ}\,
\tt_\mu \, ,\label{(128)}\fe}
in which ${\calQ}$ represents the heat loss rate per unit volume. This 
means that the energy conservation identity (\ref{(33)}) will need to be 
replaced by the energy loss formula
{\be {\calQ}+\Thet\nabla_{\!\nu}\ses^\nu=
({\calE}^{\rm c} - {\calE}^{\rm f})\nabla_{\!\nu}\nn_\rmf^{\, \nu}
+\nn_{\rm f}^{\,\nu}\varppi^{\rm f}_{\,\mu\nu}\uu^\nu = \kkappa\,(
{\calE}^{\rm c} - {\calE}^{\rm f})^2+\cc_\rmr\,\nn_\rmf\,\ww^\rmf 
\vv_\rmr^{\,2} \, ,\label{(129)}\fe}
in which the right hand side is the combination of terms that 
(by the non negativity of the coefficients $\kkappa$ and $\cc_\rmr$)
has been made to satisfy the positivity condition (\ref{(121)}).

If ${\calQ}$ is prescribed (by what would be a third secondary equation 
of state) as a function of the relevant densities -- particularly that of 
the entropy which determines the temperature $\Thet$ --  then 
(\ref{(129)}) will in principle provide what is needed for calculating 
the evolution of the entropy current. However, as the temperature 
dependence of ${\calQ}$ is likely to be highly sensitive \cite{Haensel92}, 
such an approach might not be accurate in practice. It might be more 
realistic to suppose that the temperature would adjust itself to the 
rather low roughly constant value needed to avoid accumulation of entropy, 
which incidentally is the consideration that justifies, as a reasonable
approximation, the neglect of heat conduction in this work. This
means adopting the conservation postulate (\ref{(37)}) to the effect that 
the second term on the left of (\ref{(129)}) will simply drop out. The 
implication is that $\calQ$ will be given just by the positive terms on 
the right of (\ref{(129)}), and that it will therefore vanish in the non 
dissipative limit cases that are obtainable, as described in Subsection 
\ref{pineq}, by taking infinite or zero values of  the dissipation
coefficients $\kkappa$ and $\cc_\rmr$.

\vfill\eject
\bigskip
{\bf Acknowledgements.} The authors wish to thank S. Bonazzola, N. Chamel, 
D. Langlois, P. Haensel and L. Samuelsson for helpful discussions on 
various occasions.

\end{document}